\documentclass{aastex631}

\usepackage{graphicx}
\usepackage{wrapfig}

\usepackage[utf8]{inputenc}
\usepackage{comment}

\usepackage{hyperref}
\hypersetup{
    colorlinks=true,
    linkcolor=blue,
    filecolor=magenta,      
    urlcolor=cyan,
}

\usepackage{amsmath,amssymb,amsfonts}
\usepackage{graphicx}
\usepackage{textcomp}
\usepackage{booktabs}
\usepackage{xcolor}
\usepackage{soul}
\usepackage{svg}
 \usepackage{mathtools}

\newcommand{\ip}[2]{\left\langle#1,#2\right\rangle}

\begin{document}

\title{Using Astrometry to Break Degeneracies in Stellar Surface Mapping}

\correspondingauthor{Jamila Taaki}

\author[0000-0001-5475-1975]{Jamila S. Taaki}
\affiliation{Michigan Institute for Data Science, University of Michigan\\
500 S State St, Ann Arbor, MI 48109
}
\email{xiaziyna@gmail.com}

\author[0000-0002-5466-3817]{Lia Corrales}
\affiliation{Department of Astronomy, University of Michigan\\
West Hall, 1085 S. University, Ann Arbor, MI 48109}
\email{liac@umich.edu}

\author[0000-0002-2531-9670]{Alfred O. Hero III}
\affiliation{Department of Electrical Engineering and Computer Science, University of Michigan\\
1301 Beal Avenue, Ann Arbor, MI 48109}
\email{hero@eecs.umich.edu}

\begin{abstract}
Astrometric jitter noise arises when starspots on a rotating stellar surface move in and out of view, shifting the photocenter. This noise may limit our ability to detect and weigh small, sub-Neptune-sized planets around active stars. By deriving a linear forward model for the astrometric jitter signal of a rotating star in a spherical-harmonic coordinate system, we show that jitter noise can be used to reconstruct surface-brightness maps and, in principle, disentangle jitter from stellar reflex motion. Furthermore, we show that astrometry and photometry probe complementary stellar surface information: photometry measures even-degree spherical harmonic surfaces that are symmetric about the equator, while astrometry measures odd-degree modes. Their joint use, therefore, breaks degeneracies in stellar surface mapping. Our model further quantifies the variation in the astrometric signal with inclination angle, which is foundational for studies of worst-case configurations of astrometric star-spot noise. For example, we show that pole-on stellar inclinations lead to poorly constrained inversions, as any stellar surface produces a purely circular astrometric jitter signal. We characterize the degeneracy in jointly identifying the stellar surface and stellar inclination, and develop a surface estimation approach. Using this approach, we present example simulations and reconstructions that demonstrate the use of astrometry data alongside light-curve data to improve stellar surface mapping and localize spot positions in latitude and longitude.  With forthcoming high-precision Gaia astrometry, astrometric surface mapping provides a promising new approach to probe stellar activity.

\end{abstract}

\section{INTRODUCTION} 

In a star-planet system, each body orbits the mutual center of mass; this is called reflex motion. By precisely measuring the star's orbital reflex motion, astrometry has thus far been used to detect and estimate the masses of large exoplanets and companion stars \citep{Stefansson_2025}. Detecting Earth-mass planets in the habitable zones of Sun-like stars, however, requires sub-microarcsecond sensitivity (0.3 $\mu$as for an Earth-Sun analog) \citep{exoplanet_perry}. Achieving sub-microarcsecond astrometric precision is the most promising route to mass determination of an Earth analogue, as, compared to the main alternative, radial velocity (RV), the expected signal-to-noise is an order of magnitude less impacted by stellar variability \citep{simlite}. Unlike detecting transit signatures, where the geometric likelihood of a planet transit falls off with the planets semi-major axis, an exoplanets astrometric signal strength $\alpha$ increases with semi-major axis: $\alpha \propto \frac{M_p}{M_\star} \frac{a}{D} $, where $\frac{M_p}{M_\star}$ is the planet-to-star mass ratio, $a$ is the planets semi-major axis and $D$ is distance \citep{shao2009astrometric}. Furthermore, astrometry obtains true mass estimates because it fits the sky plane orbit, unlike RV, which yields only a minimum exoplanet mass $M_p \sin(i)$ as the orbital inclination $i$ is unknown \citep{exodetect}. Diffractive pupil astrometry \citep{diff_pupil} has been proposed as a method to achieve sub-microarcsecond astrometric precision, and lab demonstrations have shown feasibility \citep{Bendek_2013}. This technique uses diffraction spikes to calibrate deformations in the optical field and obtain relative astrometry from background stars. Proposed diffractive-pupil missions include TOLIMAN \citep{toliman} to search for planets around Alpha Centauri, SHERA \citep{SHERA}, and integrated coronagraph-astrometry telescopes \citep{astro_coron}. Once sub-microarcsecond astrometric precision is reached, starspot noise is the major limiting factor in detecting and measuring exoplanets.

Starspots caused by magnetic activity can cause significant shifts in the photo-center of the star, termed \textit{astrometric jitter}, that can obscure true reflex motion \citep{sun-jitter}. Astrometric jitter poses a significant challenge for measuring Solar-system analogs, and mitigating starspot noise has been identified as a key area for future astrometry and RV missions \citep{exep}. In this work, we show that the astrometric jitter of a rotating star encodes surface-brightness information that can be used to map stellar surfaces and help constrain starspot features. While light curves have been previously used to map stellar surfaces \citep{russell, matrix-inversion, rottenbacher, cowan, cowan18, Luger_2021, Luger_2021_2}, astrometry probes complementary surface information and can better localize the positions of starspots on stellar surfaces. In principle, a joint astrometric fit could recover both the stellar surface and the true stellar reflex motion.

Starspots form where regions of strong magnetic flux suppress convection, producing cooler, darker regions on the photosphere \citep{sunspot}. While solar-like levels of starspot coverage, some $0.03 \%$ of the visible stellar surface \citep{morris2018spotting}, have limited impact on astrometric mass measurement precision \citep{catanzarite2008astrometric, sun-jitter, Lindegren_07}, younger, more active or rapidly rotating stars may far exceed the Sun’s 0.5 $\mu $as level of jitter at 10 pc, posing a major challenge for mass determination \citep{Shapiro_2021, starspot_age}.  In particular, M dwarfs often host large spot complexes that can remain stable over many rotations. For example, the M-dwarf TOI-3884 shows a polar spot with a possible $\approx 0.44 R_\star$ radius \citep{toi3884}. \citet{meunier} show that the detection of exoplanets with longer orbital periods around nearby stars will be most impacted by astrometric jitter. Furthermore, even for the best F G K candidate stars, Earth-mass uncertainties due to starspot jitter are expected to be some $30 \%$ on average \citep{meunier}. \citet{Bao_2024} utilize a single-starspot model from \citet{morris2018spotting} and simulated astrometric jitter for 78 solar-like \textit{TESS} targets, similarly finding some $30 \%$ mass uncertainty for Earth-mass planets at 10 pc. \citet{Damiano_2025} find a mass uncertainty below 10 \% is needed to identify atmospheric compositions of Earth analogues. Future missions to characterize Earth analogues, e.g., the Habitable Worlds Observatory (HWO), may also rely on planets found through astrometry. Thus, characterizing the impact of stellar signatures on astrometric measurements remains an important problem. 

The Gaia mission is set to deliver two successive data releases (DR4 and DR5, the end-of-mission release) of ultra-precise astrometric measurements for over two billion stars \citep{gaia}. With forthcoming Gaia data releases, the sensitivity required to map starspots with astrometry will soon be reached. Our work lays out the methods to do this. \citet{morris2018spotting} use an analytic spot model to show nearby ($<$5 pc) low-mass stars are excellent candidates for detecting spot-induced jitter with Gaia. Additionally, \citet{Sowmya_2022} models solar-like stars rotating faster than the sun, producing higher levels of magnetic activity, and shows that their star spot coverage is detectable with Gaia. Over its 55-month mission, Gaia will achieve $\sim 7$ $\mu$as precision for the Gaia magnitude $G < 12$ targets. For the brightest $G \sim 5 - 8$  mag nearby targets studied in \citet{morris2018spotting}, end-of-mission precisions in position/parallax of the order of $3-5$ $\mu$as are estimated after stacking $\sim 100$ visits. In anticipation of these data releases, we develop astrometric surface mapping techniques to apply to these targets. Although motivated by the need to understand starspot features and mitigate stellar jitter in astrometric surveys, more broadly, this work provides a new way to map starspots and stellar surfaces.

While Doppler imaging \citep{vogt_doppler}, and interferometry \citep{monnier_spot, Zhao_2009, rottenbacher, Parks_2021} have been used to map a handful of large, partially resolved, nearby stars, stellar surface mapping has primarily been performed using light curves, both through transit-eclipse mapping \citep{Silva_2003} and rotational light-curve mapping \citep{russell}. Light curve, or photometric, mapping uses the disc-integrated flux over time to map a stellar surface. As noted by \citet{Luger_2021}, the Kepler and TESS space telescopes, with their extreme levels of photometric sensitivity, have enabled the ensemble properties of large numbers of stars to be studied. However, light-curve inversion is fundamentally ill-posed since one seeks to reconstruct a high-dimensional surface map from low-dimensional disc-integrated flux measurements, and a unique surface map solution does not exist \citep{Walkowicz_2013, Basri_2020}. Interferometric imaging may carry more information about a stellar surface depending on the array’s baseline geometry. \citet{pope} derives analytic expressions for the interferometric visibility function response to surface spherical harmonics at a fixed stellar inclination. \citet{rottenbacher} provide a direct comparison of fitting a spot model to $\sigma$ Gem with interferometric, Doppler, and light-curve data. All three techniques reproduce the longitudinal spot distribution. However, light-curve inversion cannot reliably recover spot latitudes, and Doppler imaging also shows latitude biases. By obtaining complementary measurements, our work demonstrates that astrometry can break light curve inversion degeneracy and improve the accuracy of stellar surface maps.

\citet{russell} introduced spherical harmonics as a tool to characterize the retrievable information in the light curve of a rotating star. A pixelized basis of the stellar surface has also been used \citep{matrix-inversion}. Furthermore, analytic spot-based light-curve models have been described \citep{spot_analytic, kipping}. Spherical harmonics provide a complete orthogonal representation of any spherical surface, with higher degrees capturing progressively finer spatial structure. Working in this coordinate space \citet{russell, cowan} have derived semi-analytic expressions of disc-integrated flux (light-curve) of a rotating body over time, which are additionally used for mapping exoplanets \citep{Rauscher_2018}. \citet{Luger_2021} and \citet{cowan} both characterize the degeneracy of rotational light-curve inversion, showing that odd-degree spherical harmonics are in the null-space of the light curve measurement operator and thus are not measurable with photometry. Our work shows that astrometry does measure odd-degree spherical harmonics, completing the sampling of the spherical-harmonic basis and obtaining additional information about the stellar surface.

In this work, we explore stellar surface inversion with astrometric measurements. We derive an analytic, linear, forward model that describes the rotational astrometric signal generated by an arbitrary stellar surface at any inclination. To derive the forward model, we obtain analytic expressions for the photo-centre in terms of spherical-harmonic coefficients representing the stellar surface. Because the spherical harmonic basis is naturally ordered by spatial scale, it can be truncated at a degree set by the measurement sensitivity and the resulting spatial resolution, reducing the infinite-dimensional inversion problem to a finite-dimensional subspace. Wigner-D rotation matrices are used to obtain rotations of the stellar surface; these are semi-analytical, linear expressions for rotations in the spherical harmonic basis. Building on this, we characterize the degradation in the stellar surface information and develop a surface inversion approach. We perform proof-of-concept reconstructions of simulated stellar surfaces and demonstrate improved ability to recover stellar inclination and localize star-spot features by combining astrometry and photometry.

This paper is organized as follows. Section \ref{sec: methods} describes the forward astrometric model of a rotating stellar surface under a spherical harmonic decomposition. Section \ref{sec: results} describes the astrophysical implications of surface mapping with photometry and astrometry, and conclusions are presented in Section \ref{sec: conclusion}.

\section{FORWARD MODEL}\label{sec: methods}

\begin{deluxetable}{lll}
\tablecaption{Table of notation\label{tab: notation}}
\tabletypesize{\footnotesize}
\tablehead{
\colhead{Symbol} & \colhead{Type} & \colhead{Meaning}
}
\startdata
$h \in \{x,y\}$ & index & Measurement axis (observer-frame). \\
$l,m$ & indices & Spherical-harmonic degree ($l = 0, \dots, L$) and order ($m=-l,\dots,l$). \\
$Y_{l}^{m}(\theta,\phi)$ & function & Complex spherical harmonic. \\
$\mathbf{s}\in\mathcal{C}^{(L+1)^2}$ & vector & Stellar surface coefficients $[s_{0}^0,\; \dots, s_{l}^m, \dots,s_{L}^L]^T$. \\
$R=(\alpha,\beta,\gamma)$ & angles & Euler angles ($z$-$y$-$z$): $\alpha=0$ (sky-plane), $\beta$ = inclination, $\gamma=\omega t$ (spin). \\
$D^l(R)$ & matrix & Wigner-$D$ rotation (for $R$) matrix for spherical harmonic coefficients \\
$k^{h}_{l m}$ & scalar & First-moment (astrometric) kernel for $(l,m)$ along measurement axis $h$. \\
$\mathbf{k}^{h}_l\in\mathcal{C}^{2l+1}$ & vector & Degree-$l$ kernel over $m$ for axis $h$. \\
$A(\beta) \in \mathcal{C}^{2N \times (L+1)^2} $ & matrix & Astrometric measurement matrix (applied to $\mathbf{s}$). \\
& & $A_{1:N}(\beta) = W_\omega B_\beta^x, \; \; A_{N+1:2N}(\beta) = W_\omega B_\beta^y$ \\
$B^{h}_\beta\in\mathcal{C}^{(2L+1)\times(L+1)^2}$ & matrix &  This matrix encodes how each spherical harmonic contributes \\ & & to the astrometric signal for a measurement direction $h \in \{x, y\}$ \\ & & at inclination $\beta$. Each column corresponds to a spherical harmonic \\ & & and gives the amplitude of the measurement of $Y_l^m$.  \\
$W_\omega\in\mathcal{C}^{N\times(2L+1)}$ & matrix & Time dependent matrix for $N$ observation times. \\ & & Each column is a complex sinusoid with frequencies varying from \\ & & $-L \omega$ to $L \omega$ as $[W_\omega]_{n,m}=e^{im\omega t_n}$.\\
$\boldsymbol{\mu} \in\mathcal{C}^{2N}$ & vector & Stacked measurement $[\boldsymbol{\mu}^x, \boldsymbol{\mu}^y]^T = A(\beta) \mathbf{s}$. \\
$\mathbf{n}\sim\mathcal{N}(\mathbf{0},\sigma^2 I)$ & vector & Measurement noise with variance $\sigma^2$. \\
$\mathbf{y}$ & vector & Noisy measurements, $\mathbf{y}=\boldsymbol{\mu}+\mathbf{n}$. \\
\enddata
\end{deluxetable}
In this section, we derive a forward model that maps a rotating stellar surface to an astrometric time series. We can describe the image of a stellar surface as a linear combination of spherical harmonics, so that the astrometric time series becomes a linear function of these stellar surface weightings. This linear forward model allows us to treat surface mapping geometrically using vector-space theory. In this framework, a measured astrometric signal can be obtained by a well-defined subspace of spherical surfaces. Within this solution space, priors or regularisation can be used to choose a unique surface solution. The vector space approach to inverse problems is commonly used in signal and image processing, see \citet{bresler2000hilbert} for more details on the techniques used here. We assume a non-time-varying static surface without differential rotation. Dynamically evolving surfaces and limb darkening are left for future work.  

Uppercase symbols denote matrices (e.g., $A$), bold lowercase symbols denote column vectors (e.g., $\mathbf{s}$), and non-bolded lowercase symbols are generally scalars or functions (e.g., $s$). We denote the conjugate transpose of a matrix or vector $A$ by $A^H = (A^*)^T$ and the complex conjugate of a scalar as $s^*$. Notation is provided in Table \ref{tab: notation}.

\subsection{Spherical harmonic surface model}

Let the stellar surface brightness map of a star be defined in spherical coordinates as $s(\theta, \phi) \in \mathcal{R}$, defined for spherical polar angle $\theta \in [0, \pi]$ and azimuthal angle $\phi \in [-\pi, \pi]$. Spherical harmonics provide a complete, orthonormal basis for any spherical surface, and can therefore exactly represent any stellar surface $s(\theta, \phi)$. By expanding the stellar surface $s(\theta, \phi)$ as a linear combination of harmonics, with higher degrees capturing progressively finer structure, we can analytically characterize the rotational astrometric signal contributed by each harmonic term individually. Using these expressions, we can map an arbitrary star's surface coefficients to the measured astrometric jitter signal. 

We define the complex spherical harmonics basis set:
\begin{align}
Y_l^{m}(\theta,\phi)
= N_{l}^m \,P_l^{m}(\cos\theta)\,e^{i m \phi},
\end{align}
indexed by degree $l=0,1,\dots,L,\ \ m=-l,\dots,l$, where $P_l^m$ are the Legendre polynomials and $N_l^m$ is a constant. The spherical stellar surface can be approximated to arbitrary accuracy by its spherical harmonic expansion in this basis up to a degree \(L\) as a vector of coefficients \(\mathbf{s} \in \mathcal{C}^{(L+1)^2}\). As the maximum degree is allowed to go to infinity $L \to \infty$, the approximation becomes exact.
\begin{align} \label{eq: expansion}
s(\theta, \phi) = \sum_{l = 0}^L\sum_{m = -l}^{l} s_l^m Y_l^m (\theta, \phi)
\end{align}
Where $s_l^m = \int_0^{2\pi} \int_{-\pi/2}^{\pi / 2} s(\theta, \phi) Y_l^m(\theta, \phi) \sin(\theta) d\theta d\phi$. The spherical harmonic representation of the stellar surface is then given by $\mathbf{s} = [s_0^0, s_{1}^{-1}, s_{1}^{0}, \dots]^T$. For computation, we truncate the sum in Equation \ref{eq: expansion} to a finite value $L$. Because the astrometric sensitivity declines with increasing $l$, the maximum degree \(L\) is chosen to match the effective noise floor. This sensitivity structure is described in Section \ref{sec: kernel}.

We use the complex spherical-harmonic basis for algebraic convenience when applying rotations, but the physical surface is modeled as real by enforcing the constraint on $\mathbf{s}$ that \( s_{l}^{-m}=(-1)^m\,(s_{l}^{m})^*,\; s_{l}^0\in\mathcal{R}\) where $s^*$ is the complex conjugate of $s$. This constraint reduces the effective degrees of freedom by half (for \(m>0\)) and means the rotated surface measurements are also real. 

Astrometric measurements provide the centroid of the star, i.e., the first moment of the star normalized by the zeroth moment (disc-integrated flux) at each timestep. This is a non-linear function of the stellar surface. We consider the first moment alone, as it can be directly obtained from the instrument, and we refer to this as the astrometric signal. The astrometric operator measures the first moment of the visible hemisphere at a point in time, as projected onto the observer's 2D x-y plane. We take the observer to lie along the 3D x-axis of the spherical frame. The astrometric measurement operators are $\mathcal{A}_x : L^1(S^2) \to \mathcal{R}$ and $\mathcal{A}_y : L^1(S^2) \to \mathcal{R} $ and are given in Appendix \ref{ap: astr}, where $L^1(S^2)$ denotes the class of integrable functions defined on the unit sphere $S^2$ ($L^1$ denotes the Lebesgue function space, distinct from the maximum degree $L$) and $\mathcal{R}$ denotes the real numbers.

\begin{figure*}[ht!]
\centering
\includegraphics[width=0.5\textwidth]{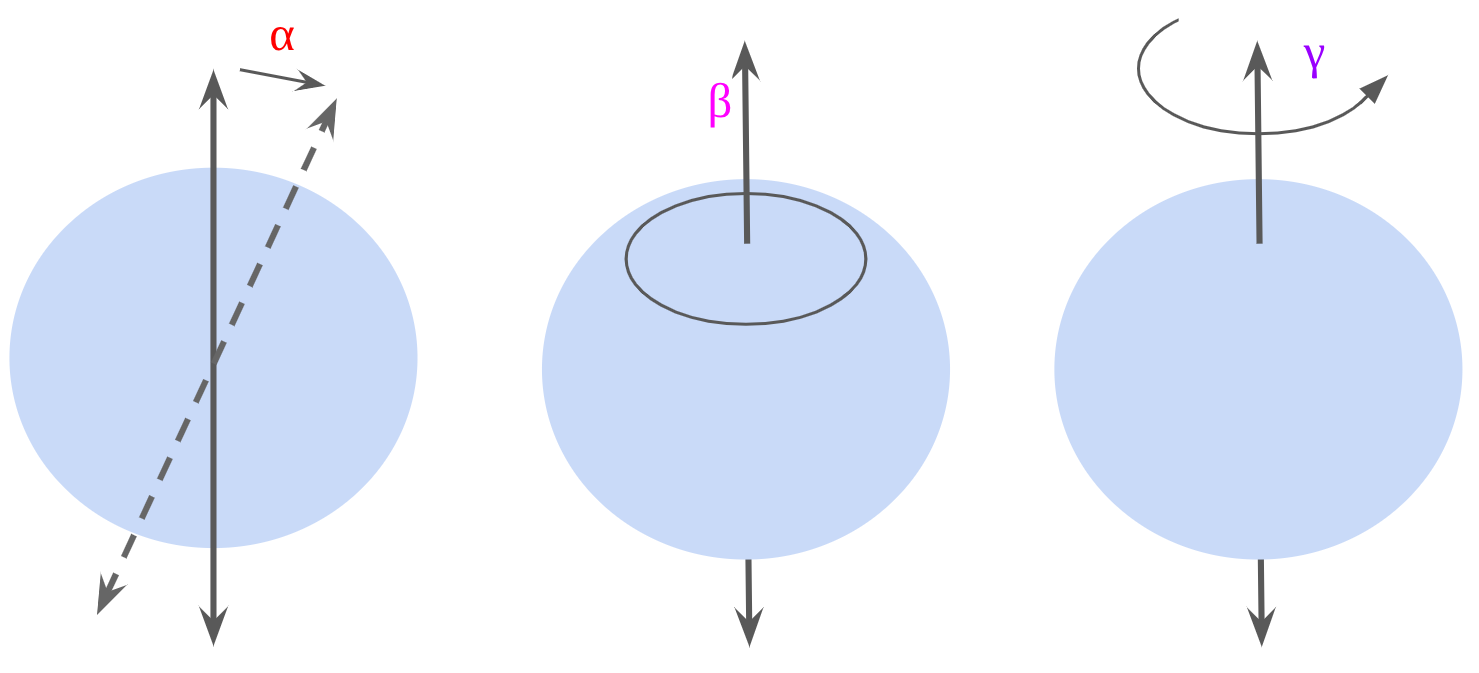}
\caption{Stellar rotation geometry (applied right to left). We parameterise the orientation with Euler angles \(R=(\alpha,\beta,\gamma)\) in the \(z-y-z\) convention, mapping the star-fixed frame to the sky via \(R=R_z(\alpha)R_y(\beta)R_z(\gamma)\). Here \(\alpha\) sets the sky-plane tilt, \(\beta\) is the inclination (tilt) of the spin axis, and \(\gamma=\omega t\) is the time-dependent rotation at angular rate \(\omega\). \label{fig: rotation}}
\end{figure*}

We use Euler angles $R = (\alpha, \beta, \gamma)$ to describe the rotation of the star about an inclined axis. Rotations are applied left to right: $\alpha$ first, followed by $\beta$ and $\gamma$ (Figure \ref{fig: rotation}). Here $\alpha$ is the sky-plane tilt and assumed to be zero. $\beta \in [0, \pi/2]$ is the inclination of the rotational axis; when $\beta = 0$, the observer faces the equator of the star, when $\beta = \pi/2$ the observer faces the pole. The time-dependent spin is $\gamma = \omega t$ with rotation rate $\omega = \frac{2 \pi}{P}$ where $P$ is the rotation period of the star.  We can represent rotations of the star directly within the spherical harmonic basis. Rotations on the sphere in the spherical harmonic basis are described by Wigner D rotations \citep{Sakurai_Napolitano_2020}, transforming the coefficient vector $ \mathbf{s}_l $ for each degree $ l $ as $ \mathbf{s}'_l = D^l(R) \mathbf{s}_l $, where $ D^l(R) $ is the Wigner D-matrix corresponding to rotation $ R $. This enables efficient rotation of the coefficient representations without direct rotations of the surfaces. 
The forward model of degree $L$ then describes the first moment $\mu^h(R): h \in \{x, y\}$ of the stellar surface after applying a rotation $R$:
\begin{align} \label{eq: astr_11}
\mu^h(R) = \mathcal{A}_h \left(\sum_{l \geq 0}^L \sum_{m = -l}^{l} [D^l(R) \mathbf{s}_l]_m Y_l^m(\theta, \phi) \right).
\end{align}
Since the astrometric operator $\mathcal{A}_h$ is a first-moment integral (defined in Appendix \ref{ap: astr}), it is a linear operator. When this linear operator is applied to a linear summation of spherical harmonics with scalar weights $[D^l(R) \mathbf{s}_l]_m$, we can equivalently take the summation of the astrometric operator applied to each static unrotated spherical harmonic as $\mathcal{A}_h(Y_l^m(\theta, \phi))$. We refer to these responses as the astrometric kernel, since they encode the sensitivity of the photocentre to individual spherical-harmonic modes.

The astrometric operator applied to spherical harmonics produces complex-valued astrometric kernel terms $k_{l,m}^h =\mathcal{A}_h(Y_l^m (\theta, \phi)) \in \mathcal{C}$, or $\mathbf{k}^h$ in vector form over all spherical harmonics and $\mathbf{k}_l^h$ in vector form over the spherical harmonics of degree $l$. This astrometric kernel is described in detail below in Section \ref{sec: kernel}. We can rewrite Equation \ref{eq: astr_11} with the astrometric kernel as:
\begin{align} \label{eq: measure}
\mu^h(R) = \sum_{l =0}^L \sum_{m = -l}^{l} [D^l(R) \mathbf{s}_l]_m  k_{l,m}^{h}.
\end{align}

Next, we note that Equation \ref{eq: measure} is an inner product between $\mathbf{k}^h_l$, the degree $l$ kernel terms, and the rotated surface coefficients $\mathbf{s}_l$. Writing the astrometric signal as an inner product, and by further utilizing the conjugate symmetry of the inner product, we obtain:

\begin{align} \label{eq: ipp}
\mu^h(R) =\sum_{l = 0}^L \ip{D^l(R) \mathbf{s}_l}{\mathbf{k}_l^h} = \sum_l \ip{\mathbf{s}_l}{D^{l, H}(R) \mathbf{k}_l^h},
\end{align}
where $D^{l,H}(R)$ is the conjugate transpose of the rotation. Because a rotation matrix is unitary, the conjugate transpose is the same as inverting the rotation, so that
\begin{align}
D^{l,H}(\alpha, \beta, \gamma) = D^l(-\gamma, -\beta, -\alpha).
\end{align} 

In this form, the astrometric measurement is given by an inner product between $\mathbf{s}$ and the rotated kernel terms.
The full forward model is a time series of photocentre measurements produced by rotating these kernel terms at different times (equivalently rotations $R = (0, \beta, \omega t)$) with the analytic Wigner-D matrices. This yields a matrix formulation that can be applied to any stellar surface.  By decomposing this matrix into time-dependent and inclination-dependent factors, and inspecting how rotations mix the kernel terms, we can cleanly characterize the information content of astrometry and photometry side by side. Below, we summarize the time-series forward model and detail the structure of the astrometric kernel to obtain our key insight that astrometry and photometry obtain distinct information about stellar surfaces.

\subsection{Time and inclination separable forward model} \label{sec: forward_mod}
For $N$ astrometric observations at times $t_1, \dots, t_N$ and $h  \in \{ x, y\}$, we define the astrometric time series $\boldsymbol{\mu}^h$ as: \begin{align}
\boldsymbol{\mu}^h = [\mu^h(R_{t_1}), \dots, \mu^h(R_{t_N})]^T.
\end{align}
In Appendix \ref{ap: rot}, a compact forward model for $\boldsymbol{\mu}^h$ is derived using Equation \ref{eq: ipp}. In this form, the measurement matrix that maps a surface to an astrometric signal is independent of the specific surface realization $\mathbf{s}$, allowing us to precompute the matrix based solely on observational parameters such as the inclination $\beta$, spin rate $\omega$, and measurement times of the following form:
\begin{align} \label{eq: main}
\boldsymbol{\mu}^h  = W_{\omega} B_\beta^h \mathbf{s}.
\end{align}
We stack the first moment time series in $x$ and $y$ and call the following description of the astrometric signal $\boldsymbol{\mu}$ the forward model:
\begin{align}\label{eq:model}
\boldsymbol{\mu}
=
\begin{bmatrix}\boldsymbol{\mu}^x\\ \boldsymbol{\mu}^y\end{bmatrix}
= A(\beta)\,\mathbf{s}, 
\qquad 
A(\beta) \;=\; 
\begin{bmatrix}
W_\omega & \\ & W_\omega
\end{bmatrix}
\begin{bmatrix}
B_\beta^x \\[2pt] B_\beta^y
\end{bmatrix}
\,. 
\end{align}
The measurement matrix $A(\beta)$ that acts on $\mathbf{s}$ encapsulates the time-dependent mapping from the visible surface to photocentre measurements. In this form, the forward model consists of a time-dependent matrix $W_{\omega} \in \mathcal{C}^{N \times (2L +1)}$ and an inclination dependent matrix $B_\beta^h \in \mathcal{C}^{(2L + 1) \times (L^2 +1)}$.  The entries of $W_\omega$ are complex sinusoids evaluated at the observation times:
\begin{align} \label{eq: W_omega}
[W_\omega]_{n,m} = e^{im\omega t_n}, \quad m = -L, \dots, L,
\end{align}
so that each column is a sinusoid of frequency $m\omega$ sampled at times $t_1, \ldots, t_N$. The astrometric signal is therefore a linear combination of complex sinusoids with frequencies at integer multiples of the spin rate, up to a maximum frequency $L\omega$. The vector $B_\beta^h \mathbf{s}$ gives the complex amplitude of each sinusoidal component, while $W_\omega$ synthesizes these amplitudes into the observed time series. The matrix $B_\beta^h$ is block-diagonal in degree $l$:

\begin{align} \label{eq: betas}
B_\beta^h = \left[\operatorname{diag}(d^0(-\beta)\, \mathbf{k}_0^h), \;\dots\; \operatorname{diag}(d^L(-\beta)\, \mathbf{k}_L^h)\right],
\end{align}
where $d^l(-\beta)$ is the small Wigner-d matrix evaluated at $-\beta$. Due to the block-diagonal form of $B_\beta^h$, each surface spherical harmonic $s_l^m$ maps to a single complex sinusoid. This mapping is determined by $B_\beta^h$ which depends on the inclination through $d^l(-\beta)$ and the astrometric kernel $\mathbf{k}$. This forward model describes observations for any inclination $\beta \in [0, \pi/2]$ and $W_{\omega}$ can be calculated for arbitrary observation times.
This decoupling of the spin (via $W_{\omega}$) and inclination (via $B_{\beta}^h$) in the forward model provides key benefits for analytical insight and computational efficiency in the inverse problem of stellar surface reconstruction. Analytically, it isolates the impact of inclination on the model, making it simple to analyse how the surface solution set changes with inclination.

\begin{figure*}[ht!]
\centering
\includegraphics[width=.5\linewidth]{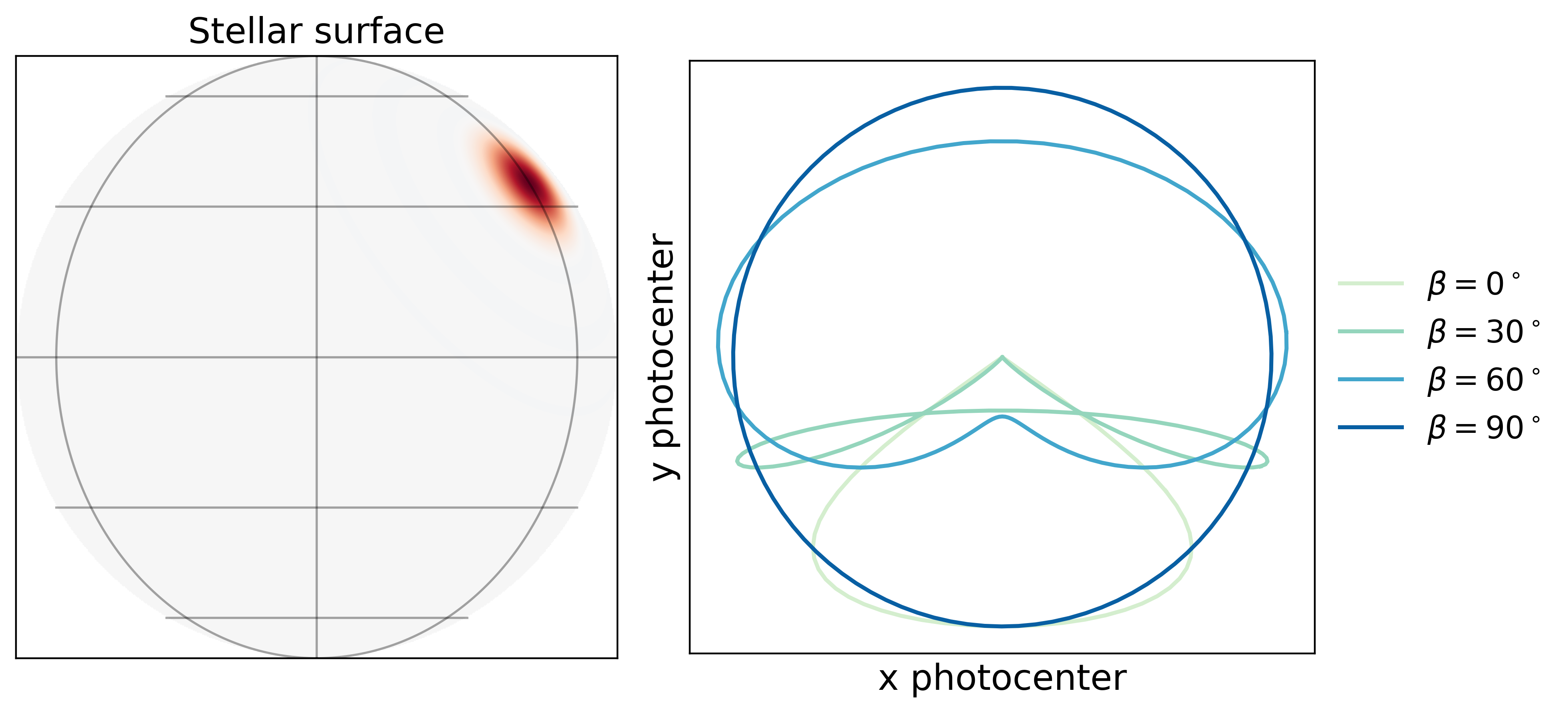}
\caption{An order $L=25$ expansion of an 8 \% $R_\star$ starspot and the $L=25$ astrometric signal. The star surface is shown (left) viewed equatorially. The astrometric signal is shown for different inclinations ranging from $\beta = 0$ (equator on) to $\beta = \pi / 2$ (pole-on). When the photo-center is at origin, the starspot is out of view. For a pole-on observer, any astrometric signal is circularized.
\label{fig: starspot}}
\end{figure*}

\begin{figure*}[ht!]
\centering
\includegraphics[width=.5\linewidth]{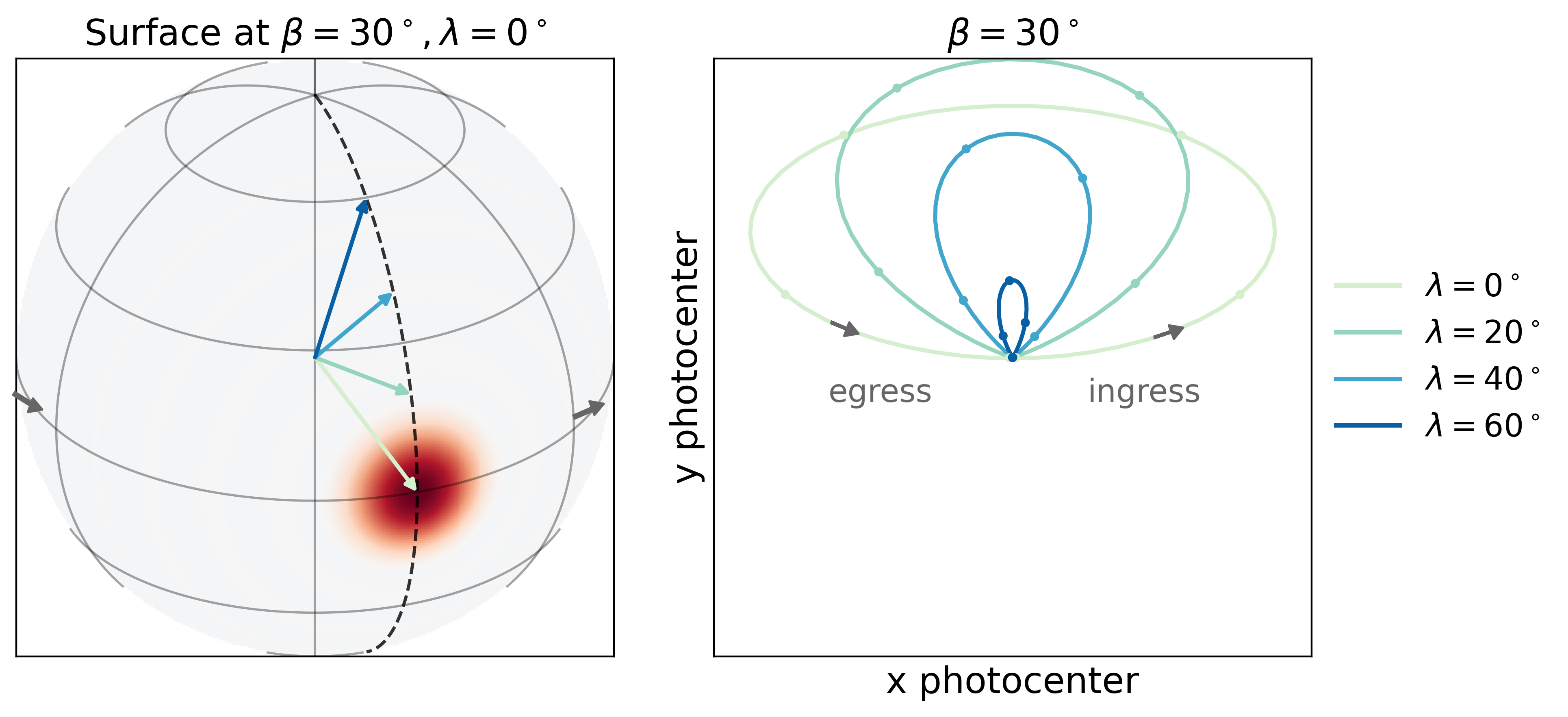}
\caption{An order $L=25$ expansion of a 10 \% $R_\star$ starspot and the $L=25$ astrometric signal. The star surface is shown (left) at an inclination of $\beta = 30^\circ$, with an equatorial starspot. For this stellar inclination, the astrometric signal from starspots at different latitudes denoted by $\lambda$ is shown (right), ranging from $\lambda = 0^\circ$ (equatorial) to $\lambda = 60^\circ$. Grey arrows denote the egress and ingress of the starspot. As the starspot heads to egress, the photocenter moves toward the origin. While the starspot is out of view, the photocenter is fixed at zero. When the starspot moves into ingress, the photocenter begins to move again. These varying starspot latitudes $\lambda$ are shown on the stellar surface (right) as arrows pointing from the star's interior origin point. Furthermore, starspots at higher latitudes generally produce less photo-center variation.
\label{fig: starspot2}}
\end{figure*}

Figures \ref{fig: starspot} and \ref{fig: starspot2} illustrate the astrometric signal for simulated starspots. Figure \ref{fig: starspot} shows how the signal shape changes with stellar inclination: for an equator-on star ($\beta = 0$), the photocenter traces a back-and-forth displacement along a single axis, while for a pole-on star ($\beta = \pi/2$), any astrometric signal is purely circular regardless of spot position. Figure~\ref{fig: starspot2} shows how spot latitude affects the signal at a fixed inclination of $\beta = 30^\circ$: spots at higher latitudes generally produce smaller photocenter displacements, and the ingress and egress of the starspot into and out of view are visible as the signal returns to zero. In both cases, the astrometric signals are non-trivial functions of inclination and spot latitude. In Section \ref{sec: kernel} we describe the astrometric sensitivity to specific spherical harmonics and in Section \ref{sec: nullspace} we describe how the recoverable surface information varies with stellar inclination.

\subsection{Astrometry measures odd degree $l$ harmonics} \label{sec: kernel}
Above, we summarized the forward model that maps a stellar surface in spherical harmonics to an astrometric time series. This forward model 
depends on kernel terms $k^h_{l,m} \; : \; h \in \{x,y\}$, which describe the astrometric sensitivity to individual spherical harmonics of degree $l$ and order $m$. In this section, we describe the kernel form and how stellar inclination affects this sensitivity. Later, in Section \ref{sec: nullspace}, we discuss the inversion degeneracy.

In a static unrotated frame, the kernel terms describe the photo-center of each spherical harmonic $k^h_{l,m} = \mathcal{A}_h(Y_l^m(\theta, \phi))$.  For our chosen convention, the astrometric kernel terms $\mathbf{k}^h : h \in \{x,y\}$ are defined for an observer along the 3D x-axis (in the spherical frame), as projected onto their 2D x-y plane of the sky (the x-y observer axis correspond to the y-z axis in the spherical frame). These kernel terms are derived as analytic integrals in Appendix \ref{ap: astr}, where we identify spherical harmonics that produce zero-valued kernel terms using the symmetry properties of the functions being integrated. 

An analogous forward model to Equation \ref{eq: main} describes photometric (flux) time series. Photometry measures the disk-integrated brightness, the zeroth moment of the visible surface, rather than the first-moment as is measured by astrometry. The photometric kernel $\mathbf{k}^{\mathrm{phot}}_l$ replaces $\mathbf{k}^h$ in $B_\beta$, while $W_\omega$ is unchanged. The photometric kernel is based on the formula for thermal light curves in \citet{cowan} that describe the flux time-series of rotating real spherical harmonics viewed equatorially, the kernel is derived in Appendix~\ref{ap: photo} for complex-spherical harmonics. 

\begin{figure*}[ht!]
\plotone{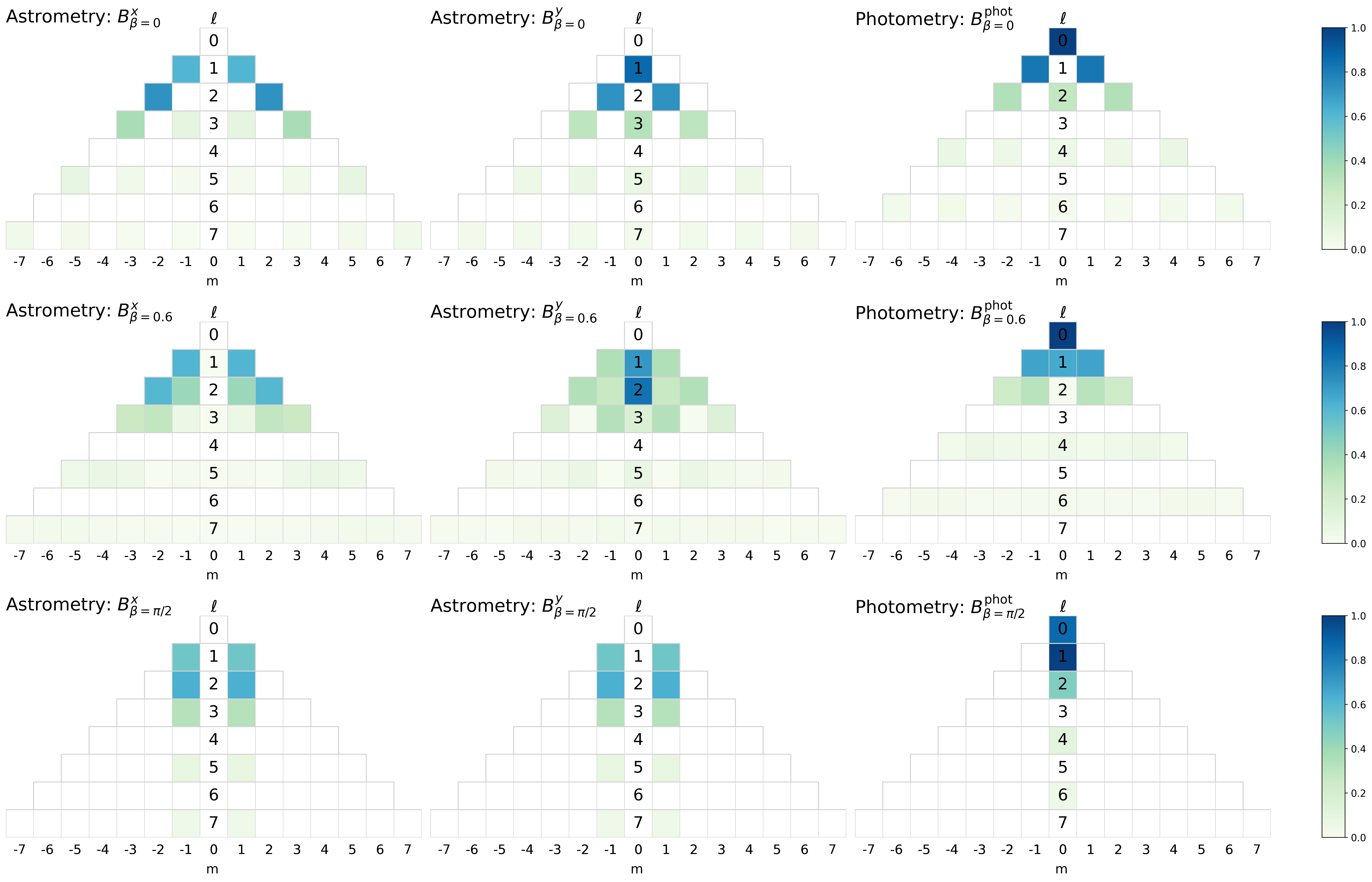}
\caption{Visualizing the inclination dependent matrix in the forward model, $B_\beta$ that describes the sensitivity to individual spherical harmonics indexed by $(l,m)$ for astrometry and photometry, for three inclinations $\beta$. 
\textit{Top ($\beta = 0$, equator-on):} For an equator-on perspective, the entries of $B_{\beta = 0}$ directly consist of the measurement kernel weights $k_{l,m}^h$. The magnitude of each kernel term describes the magnitude of photocenter shift for the corresponding $l,m$ spherical harmonic over a rotation, and for photometry the change in flux. Where the kernel terms are 0 (white), the ($l, m$) spherical harmonic produces no measurable signal throughout a full rotation. 
\textit{Middle ($\beta = 0.6$, inclined):} For an inclined star, the kernel terms $\mathbf{k}_l^h$ have been 'mixed' over order $m$ per degree $l$ by the small wigner-d matrices to form the measurement matrix $B_\beta$, rendering unobservable terms observable and increasing the recoverable surface information. 
\textit{Bottom ($\beta = \pi/2$, pole-on):} A pole-on observer retrieves the least information about a stellar surface. Only harmonics with $|m| = 1$ produce non-zero astrometric signals in both the x and y directions, resulting in a purely circular photocenter motion for any stellar surface. Similarly, photometry can only measure $m = 0$ harmonics and there will be no change in flux over a rotation.
\label{fig: general}}
\end{figure*}

A key analytic result of this work is that astrometry measures odd $l$ spherical harmonics that are anti-symmetric about the equator, in contrast to photometry, which measures even $l$ degree harmonics that are symmetric about the equator \citep{cowan}. Astrometry and photometry, therefore, probe orthogonal subspaces of the stellar surface, meaning that joint astrometry–photometry measurements can recover surface structure inaccessible to either measurement type alone. For astrometry, the kernel terms are non-zero for odd $l$ and $l = 2$, and the x-direction kernel is non-zero for harmonics of odd order $m>2$, while the y-direction kernel is non-zero for harmonics of even order $m>2$. For these non-zero terms, the astrometric kernel weights are integrals over the Legendre polynomials $P_l^m(u)$ given by:

\begin{equation} \label{eq: k_summary}
\begin{aligned}
k_{l,m}^x &=
\begin{cases}
\frac{I_{\phi, x}(m) N_{l}^{m}}{\pi}\displaystyle\int_{-1}^{1} (1-u^{2})\,P_{l}^{m}(u)\,d u
  & \text{if } (l \text{ odd} \;\text{and}\; m \text{ odd}) \;\text{or}\; (l=2 \;\text{and}\; |m|=2).\\
0 & \text{otherwise,}
\end{cases}
\\[6pt]
k_{l,m}^{y} &=
\begin{cases}
\frac{I_{\phi, y}(m) N_{l}^{m}}{\pi}\displaystyle\int_{-1}^{1} u\sqrt{1-u^{2}}\,P_{l}^{m}(u)\,du
  & \text{if } (l \text{ odd} \; \text{and} \; m \text{ even}) \;\text{or}\; (l=2 \;\text{and}\; |m|=1),\\
0 & \text{otherwise.}
\end{cases}
\end{aligned}
\end{equation}
Where $u$ is the variable of integration and $I_{\phi, h}(m)$ terms are given in Appendix \ref{ap: astr}. We evaluate the inner integral numerically using Sympy \citep{symp}, achieving a relative error of order $\sim 10^{-15}$. The kernel terms need only be evaluated once and can be used to derive the rotational signatures for different stars or different inclinations. These are shown evaluated up to $L = 7$ for an equatorial $\beta = 0$ stellar inclination in Figure \ref{fig: general}, alongside the photometry kernel derived by \citet{cowan}. For an equatorial observer ($\beta = 0$), the astrometric kernel exactly describes the relative weighting of spherical harmonics to the astrometric signal because the Wigner-D matrix in Equation \ref{eq: ipp} is an identity matrix. The kernel terms in Equation \ref{eq: k_summary} illustrate that astrometric measurements of a star viewed equatorially ($\beta = 0$) probe independent spherical harmonics; the x direction probes odd-order harmonics, while the y direction probes even-order harmonics. Additionally as special cases, the x direction is sensitive to $l=2, m \pm 2$, and the y direction to $l=2, m= \pm 1$. These are the only even-degree terms accessible to astrometry. Conversely, photometry also measures $l=1, m = \pm 1$ as a special case.  Intuitively, the astrometric sensitivity pattern shown in Figure \ref{fig: general} arises because the astrometric operator measures the visible centroid, which integrates an antisymmetric position weighting multiplied by the visible surface. Hence, this will only be non-zero if the visible surface is asymmetric. In contrast, because the photometry kernel measures the summed visible flux which is a symmetric weighting of the surface, the photometric kernel is non-zero only for equatorially symmetric harmonics.

When a star is inclined, the kernel terms for each degree $l$ are mixed among orders \(m \in \{ -l \dots l \} \) within each \(l\) as $(\mathbf{k}^h_{l})' = d^l(-\beta)\, \mathbf{k}^h_l$, where $d^l(-\beta)$ is the small Wigner-d matrix. The matrix $B_\beta^h$, which is derived in Appendix \ref{ap: rot}, captures this and is visualized in Figure \ref{fig: general}. Because mixing is applied separately per degree $l$, even-degree $l > 2$ harmonics with zero-valued kernel terms remain zero after mixing and are unobservable. Likewise, for photometry, odd-degree $l > 1$ harmonics remain unobservable. Consequently, astrometry and photometry remain complementary when the star is inclined.

However, as described in \citet{cowan}, because of this mixing by $d^l(-\beta)$, for an odd-degree $l$, all order $m$ reweighted kernel terms in the vector $(\mathbf{k}^h_{l})'$ are non-zero. Thus, the corresponding spherical harmonics are observable when the star is at an interior inclination $\beta \in ( 0, \pi/2 )$. As a result, a larger number of spherical harmonics contribute to the astrometric or photometric signal when the star is inclined, and we retrieve more information about the stellar surface. We discuss the implications of this result for mapping the surface below in Section \ref{sec: nullspace}. 

For the polar observer $\beta = \pi/2$, all but the $m = \pm 1$ harmonics are weighted to zero. Figure \ref{fig: general} shows the kernel values for this case: only the $|m|=1$ column is non-zero, and the x and y kernels are identical up to a sign, reflecting the circular symmetry of the signal.

As can also be seen in Figure \ref{fig: general}, the individual coefficients $k^h_{l,m}$ decrease in magnitude with increasing degree $l$. This diminishing sensitivity allows us to truncate the spherical harmonic representation at a maximum degree $L$ while maintaining a desired level of accuracy. However, this does not necessarily imply that higher-degree terms are negligible in the total observed signal. Each degree $l$ contains $2l+1$ spherical harmonic modes, so the aggregate contribution of terms above a given degree $l$ depends on the rate of decay of $|k^h_{l,m}|$ and the growing number of modes. In other words, even if each coefficient becomes small, the sum over all degrees and orders may be a significant factor in the spherical harmonic approximation. 

Below, we summarize the astrometric signal induced by individual spherical harmonics $Y_l^m (\theta, \phi)$ for the extreme cases of inclinations $\beta = 0$ (equator on) and $\beta = \pi/2$ (pole-on).

\paragraph{Rotating star viewed equatorially: $\beta = 0$}
In the equatorial case, the astrometric kernel terms $\mathbf{k}^h$ directly weight complex exponential oscillation terms, we obtain the following astrometric measurements for the stellar surface harmonics:
\begin{align} \label{eq: face_on}
\mu^{x}_{l,m}(t) =
\begin{cases} 
 e^{im\omega t} k^{x}_{l,m} & \text{if } (l \; \text{odd} \; \text{and} \; m \; \text{odd}) \;\text{or}\; (l=2 \;\text{and}\; |m|=2),\\
0 & \text{otherwise},
\end{cases} \\ \label{eq: face_on_y}
\mu^{y}_{l,m}(t) =
\begin{cases} 
e^{im\omega t} k^{y}_{l,m} & \text{if } 
(l \text{ odd} \; \text{and} \; m \text{ even}) \;\text{or}\; (l=2 \;\text{and}\; |m|=1),\\
0 & \text{otherwise},
\end{cases}
\end{align}

\paragraph{Rotating star viewed from the pole: $\beta = \pi / 2$}
This case is the most ill-conditioned, with the least information about the stellar surface. For any stellar surface, the signal is purely circular. Equivalently, the astrometric harmonics are non-zero only for $|m| = 1$. The terms $p_{l,m}^h$ are weightings defined in Appendix \ref{ap: pole}

\begin{align}
\mu^x_{l ,m} (t) &=
\begin{cases}  \label{eq: pole_x}
e^{i m \omega t} p^x_{l,m}
& \text{if } (l \text{ odd} \; \text{or} \; l=2) \; \text{and} \; |m|=1 \\
0 & \text{otherwise},
\end{cases} \\ \label{eq: pole_y}
\mu^y_{l,m} (t) &=
\begin{cases}
e^{i m \omega t} p^y_{l,m}
& \text{if } (l \text{ odd} \; \text{or} \; l=2) \; \text{and} \; |m|=1 \\
0 & \text{otherwise},
\end{cases}
\end{align}

Finally, since the observed first moment is found as the sum over the contributions in $l,m$, for a stellar surface described by coefficients $s_l^m$, the overall astrometric signal is given by:
\begin{align}
\mu^x (t) = \mu^y(t) \propto e^{i \omega t} \sum_{odd \;  l} (s_l^1 + s_l^{-1}).
\end{align}

Although this is the most weakly constrained case for inversion, it can also produce the strongest signal as the rotated kernel weightings $p_{l,m}^h$ are large and produce equal displacements in both x and y directions. Consistent behaviour is seen by \citet{Sowmya_2021}, showing pole-on configurations produce significant astrometric jitter that can interfere with measurements of planetary reflex motion.

\subsection{Astrometric inversion degeneracy} \label{sec: nullspace}
\subsubsection{Rank and nullity}
\begin{figure*}[ht!]
    \centering
    \includegraphics[width=.5\linewidth]{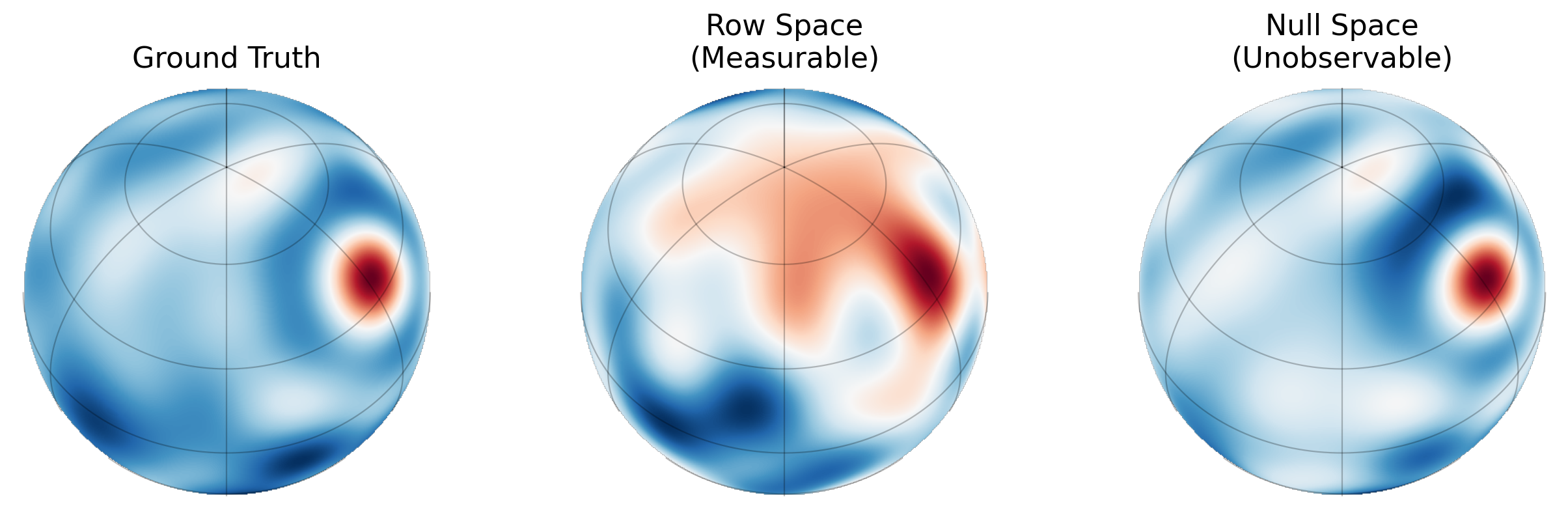}
    \caption{Decomposition of a stellar surface (left) into its measurable row-space (center) and unobservable null-space (right) components under the astrometric forward model at $\beta = 0.91$ and $L = 10$. For this configuration, 33\% of the surface energy lies in the row-space and 67\% in the null-space. Reconstruction examples for the astrometric measurements of this surface are shown below in Figure \ref{fig: recon}. 
    \label{fig: decomposition}}
\end{figure*}

Although many spherical harmonics can produce non-zero astrometric signals, subsets of these observable harmonics can produce indistinguishable time series, making the inversion degenerate. The row-space and its complement, the null-space of the astrometric operator $A(\beta)$, characterize the space of possible stellar surface solutions $\mathbf{s}$ for noiseless measurements $\boldsymbol{\mu}$ and how this varies with inclination angle $\beta$. The null-space denoted as $N(A(\beta))$ consists of all surfaces $\mathbf{s} \in N(A(\beta))$ such that $A(\beta) \mathbf{s} = 0$, meaning these components produce no detectable astrometric motion in $x$ or $y$ and are thus inherently unobservable. Conversely, the row space of $A(\beta)$ describes all stellar surface coefficient vectors $\mathbf{s}$ that produce non-zero measurements and have no component lying in the null space. The row space is the orthogonal complement of the null-space $\mathrm{Row}(A(\beta)) = \mathcal{N}(A(\beta))^\perp$, thus any surface can be uniquely decomposed into a combination of a row space vector and a null space vector. An example decomposition for a simulated stellar surface is shown in Figure \ref{fig: decomposition}.
\[
\mathbf{s} = \mathbf{s}_{Row(A)} + \mathbf{s}_{N(A)}
\]

If a particular surface $\mathbf{s}_{Row(A)}$ produces an astrometric signal $\boldsymbol{\mu}$, then any additive surface drawn from the null space $\mathbf{s}_{N(A)}$ leaves the astrometric signal unchanged:
\[
A(\beta)\,\mathbf{s}_{Row(A)} = \boldsymbol{\mu} \quad \Rightarrow \quad
A(\beta)\,\big(\mathbf{s}_{Row(A)} + \mathbf{s}_{N(A)}\big) = \boldsymbol{\mu}
\quad \text{for all } \mathbf{s}_{N(A)} \in N\big(A(\beta)\big).
\]
Given an astrometric signal $\boldsymbol{\mu}$, solving for a surface $\mathbf{s}$ is unique only up to the row space. Then, the size of the null space characterizes the bias in estimating a stellar surface from measurements. If the null space spans all spherical harmonics, it would mean we could not identify any surface solution. If the null space is empty, we can uniquely constrain the stellar surface from the measurements. The dimension of the null space is denoted as $\operatorname{nullity}(A(\beta))$ and is thus a measure of the surface identifiability. The larger this nullity is, the less well constrained the inversion is, as a larger set of surfaces can produce the observations. Briefly, the dimension of a vector subspace is the number of linearly independent vectors needed to represent every vector in that subspace by linear combination. For example, in a two-dimensional Cartesian space, every point can be written as a combination of two basis vectors, $x$ and $y$. 

To understand what populates the null space of $A(\beta)$, we distinguish between two types of unobservable surfaces. The null space can contain both individual spherical harmonics that produce no astrometric signal and specific linear combinations of spherical harmonics that are individually observable. If an individual spherical harmonic $Y_l^m$ is in the null space for inclination $\beta$ :
\[
A(\beta)\,\mathbf{e}_{l,m} = \mathbf{0},
\]
Here $\mathbf{e}_{l,m}$ is the canonical basis vector with a $1$ in the position corresponding to $(l,m)$ and zeros elsewhere, representing a surface consisting solely of the $(l,m)$ spherical harmonic mode. Other null space directions correspond to linear combinations of otherwise measurable harmonics whose contributions to the astrometric signal cancel exactly. To see how such degeneracies arise, consider the case $\beta = 0$ (equator-on) described in Equation \ref{eq: face_on}. All spherical harmonics $Y_l^m$ of the same order $m$ produce the same time-varying signal of the form $e^{im\omega t}$, weighted by different scalar kernel terms $k_{l,m}^h$. Thus, a surface consisting of a linear combination of order $m$ harmonics can be constructed where the astrometric signal is zero even though the elements of $\mathbf{s}$ are nonzero.

Instead of deriving the nullity by explicitly constructing a null space for each inclination, we obtain the nullity of $A(\beta)$ directly from its rank via the rank–nullity theorem. The rank-nullity theorem states that for a linear matrix transform \( A(\beta) : \mathcal{C}^{(L+1)^2} \to \mathcal{C}^{N} \),
\begin{align} \label{eq: rank_null}
\operatorname{nullity}(A(\beta)) + \operatorname{rank}(A(\beta)) = (L+1)^2,
\end{align}
Here, we assume the number of measurements $N > (L+1)^2$, where $L$ is the maximum spherical harmonic degree. The $\operatorname{nullity}(A)$ is the number of linearly-independent vectors $\mathbf{s} \in \mathcal{C}^{(L+1)^2}$ that are in the null-space of $A(\beta)$: $A(\beta)\mathbf{s} = 0$. Similarly, the $ \operatorname{rank}(A(\beta))$ is the number of linearly independent columns or rows of $A(\beta)$. The rank represents the number of independently recoverable surface degrees of freedom. The dimension of the domain (i.e., the total number of spherical harmonic terms) is $(L+1)^2$.

\begin{figure*}[ht!]
\plotone{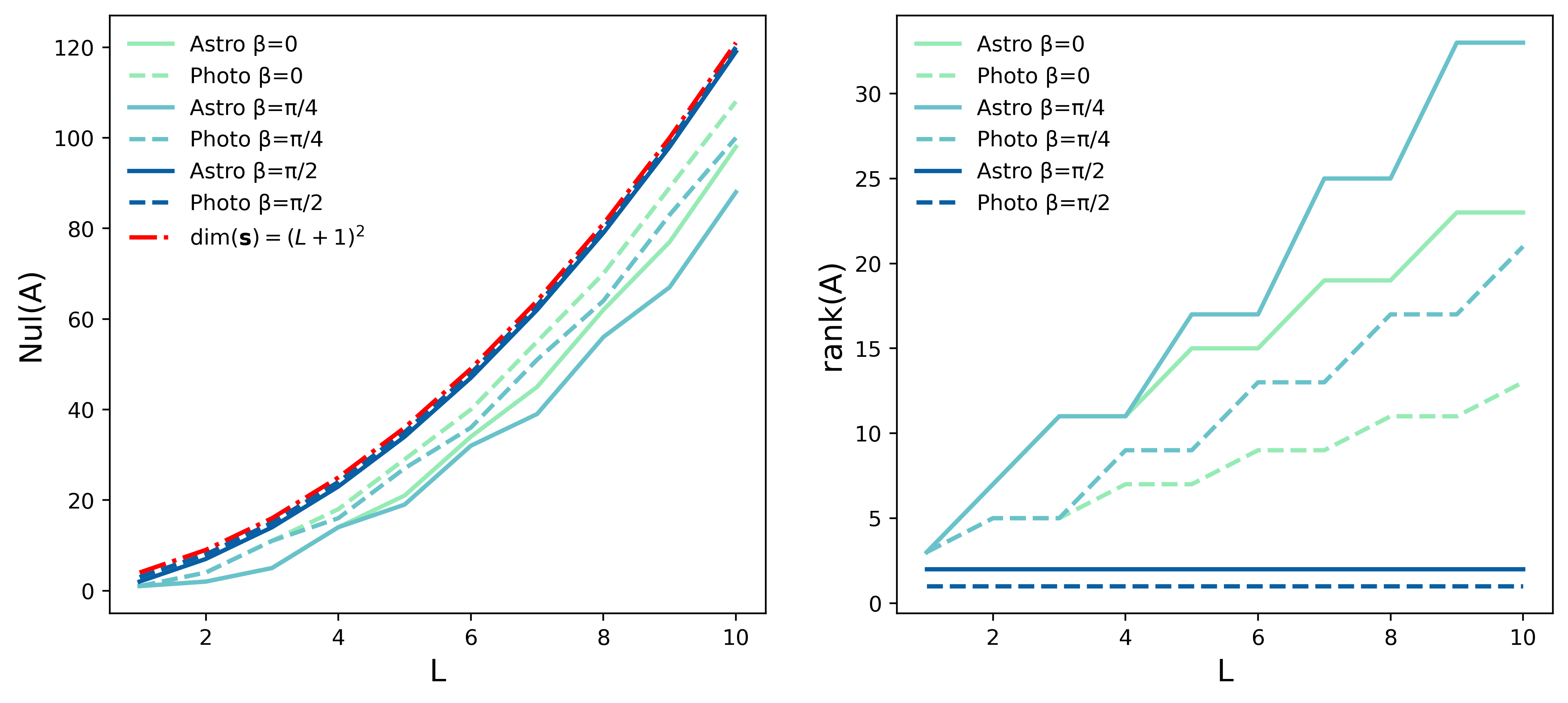}
\caption{Comparing astrometry and photometry null space and rank across different inclinations. The size of the nullspace (left) is bounded by the rank of the measurement (right) subtracted from the total number of spherical harmonics $(L+1)^2$ (red line). The rank quantifies the total information in the astrometric or photometric measurement. Astrometry has a higher rank than photometry since it has two measurement directions; therefore, more information about the surface can be obtained. In general, both operators have large null spaces relative to the total number of spherical harmonics. With increasing $L$, the size of the null-space approaches the total number of spherical harmonics $(L+1)^2$,  producing more degraded surface inversions. When astrometry and photometry observations are combined, the combined rank is obtained by adding the individual rank curves (right), see Table \ref{tab: rank}. Thus, the combined null space is smaller than either individual null space. 
\label{fig: rank}}
\end{figure*}

The rank of $A(\beta)$ for an order $L$ spherical harmonic expansion is summarized in Table \ref{tab: rank} for astrometry, photometry, and both measurements combined. Furthermore, in Figure \ref{fig: rank}, the nullity and rank of the forward model are shown for varying $L$ and $\beta$. There are several key implications for surface mapping obtained from these results. First, the null space for photometry is larger than the null space for astrometry at all inclinations, meaning that photometric inversion carries less surface reconstruction power. Conversely, the rank of the photometry model is lower than that of the astrometry model. This is expected, since astrometry measurements have two degrees of freedom, whereas photometry measurements have only one. A higher rank means the null space is smaller and thus more information about the stellar surface can be gleaned. For both astrometry and photometry, the null-space is smallest at interior inclinations between 0 and $\pi/2$ and largest at the pole. Second, since the dimension of $\operatorname{dim}(\mathbf{s}) = (L+1)^2$ grows quadratically, while the model rank $\propto L$ grows linearly for both photometry and astrometry, at a higher degree $L$, small spatial scale surface information is unrecoverable. This result was previously shown for photometry in \citet{Luger_2021}. Below, we derive these results in detail.

Since \( A(\beta) = W_\omega B_\beta \), the rank of \( A(\beta) \) is bounded as:
\begin{align} \label{eq: rank_bound}
\operatorname{rank}(A(\beta)) \leq \min(\operatorname{rank}(W_\omega), \operatorname{rank}(B_\beta)).
\end{align}

The matrix \( B_\beta^h \in \mathcal{C}^{(2L+1) \times (L+1)^2} \) has at most $2L+1$ rows, so \(\operatorname{rank}(B_\beta^h) \leq (2L+1)\). Since the combined matrix $B_\beta = [B_\beta^x;, B_\beta^y]$ stacks both directions and has $2(2L+1) = 4L+2$ rows, $\operatorname{rank}(B_\beta) \leq 4L+2$. Looking at the dimensions of \( W_\omega \in \mathcal{C}^{N \times (2L+1)} \), the overall model rank in Equation \ref{eq: rank_bound} is bounded as $\operatorname{rank}(A(\beta)) \leq \min (N, 2L+1)$. Taking $N\!\ge\!2L{+}1$ measurement times with uniform time-sampling, the complex sinusoids that form the columns of $W_\omega$ are mutually orthogonal \citep{oppenheim1993digital}, so $W_\omega$ is full rank ($2L+1$). This means that as long as $N\!\ge\!2L{+}1$, the rank of $A(\beta)$ is solely determined by the rank of $B_\beta$. The rank of \( B_\beta \) depends on the inclination \(\beta\), we explore the different cases below.  Note, the nullity must be derived jointly for $x$ and $y$ unless $B^x_\beta$ and $B^y_\beta$ are linearly independent, which is the case when $\beta = 0$ but not guaranteed for other inclinations. 

Following from the kernel description in Section \ref{sec: kernel} and the Equations \ref{eq: face_on} and \ref{eq: face_on_y}, for $\beta = 0$ (equator-on), the null space includes all even degree-$l$ harmonics with $l>2$, as well as specific linear combinations of harmonics that share the same azimuthal order $m$ but differ in degree $l$. Among the azimuthal indices $m \in [-L,L]$, the number of odd values of $m$ is $2\lceil L/2\rceil$ and the number of even values is $2\lfloor L/2\rfloor + 1$. In addition, $\mu^x$ is non-zero for two extra harmonics with $(l,m) = (2,\pm 2)$, while $\mu^y$ measures two extra harmonics with $(l,m) = (2,\pm 1)$. Taken together, this yields $\operatorname{rank}\big(A(\beta=0)\big) = 2L + 5$. Applying the rank–nullity theorem in Equation~\ref{eq: rank_null}, the null space therefore has dimension $\operatorname{nullity}\big(A(\beta=0)\big) = (L+1)^2 - (2L+5) = L^2 - 4$.

For general inclinations $\beta \in (0, \pi/2)$, the Wigner-d rotation mixes harmonics across azimuthal orders, which changes the null space structure considerably. For inclinations $\beta \in (0,\pi/2)$, the inclination–dependent matrices, $B^h_\beta$ for $h \in \{x,y\}$, have rank 2L + 1. In Equation \ref{eq: mix}, these matrices are assembled from blocks $d^l(-\beta)$ that mix the astrometric kernel terms $\mathbf{k}^h_l$ across $m \in \{-l,\dots,l\}$. Each $d^l(-\beta)$ is a small Wigner-d matrix of full rank. As a result, for inclined stars, all azimuthal orders $m \in \{-l,\dots,l\}$ for odd degrees $l$ contribute to the observed signal. The full forward matrix $A(\beta) = W_\omega B_\beta$ has $\operatorname{rank}(A(\beta)) = 4L - 2$. This is lower than the sum of the ranks of the $x$ and $y$ components because spherical harmonics of the degree $l = L$, order $m = \pm L$ and $m = \pm (L-1)$ spherical harmonics produce identical $x$ and $y$ measurements up to a scaling, reducing the rank by $4$. For $\beta = \pi/2$ (pole-on) by inspection of Equation \ref{eq: pole_x} and \ref{eq: pole_y}, both directions produce only a single $m=\pm1$ non-zero harmonic and $\operatorname{rank}(A(\beta = \pi/2)) = 2$.

\begin{table}[h]
\centering
\caption{Rank of the Forward Model $\operatorname{rank}(A(\beta))$ ($L>2$). } 
\begin{tabular}{lllc}
\toprule
Inclination (\(\beta\)) & $\operatorname{rank}(A_{astro}(\beta))$ & $\operatorname{rank}(A_{photo}(\beta))$ & $\operatorname{rank}(A_{astro \; \cap \; photo}(\beta))$ \\
\midrule
\(\beta = 0\) (Face-on) & 
    $2L + 5$ & $L + 2$ & $3L + 7$ \\
\(\beta \in (0, \pi/2)\) (Inclined) & 
   $4L - 2$ & $2L - 1$ & $6L - 3$ \\
\(\beta = \pi/2\) (Pole-on) & 
    $2$ &  $1$ & $3$ \\
\bottomrule
\end{tabular}
\tablecomments{The size of the null space is $(L+1)^2 - \operatorname{rank}(A(\beta))$.}
\label{tab: rank}
\end{table}

\subsubsection{Inclination identifiablity}

Above, we considered the surface inversion bias for a given inclination $\beta$. In this section, we consider estimating both the stellar surface $\mathbf{s}$ and stellar inclination $\beta$ jointly. Without additional constraints, the inclination cannot be uniquely identified, and consequently, neither can the stellar surface. In the forward model, an astrometric observation $\boldsymbol{\mu}$ is a linear combination of the columns of $A(\beta)$ scaled by the surface spherical harmonic coefficient elements of $\mathbf{s}$. The column space describes the set of all possible astrometric signals at a fixed inclination:
\begin{align}
\operatorname{col}(A(\beta)) = \{A(\beta) \mathbf{s} \; : \; \mathbf{s} \in \mathcal{C}^{(L+1)^2}\}.
\end{align} If this set is invariant with inclination, then at any interior inclination $\beta \in (0, \pi/2)$ there exists a surface $\mathbf{\hat{s}}_\beta$ that can produce the same astrometric signal $\boldsymbol{\mu}$:
\begin{align}
\boldsymbol{\mu} = A(\beta) \hat{\mathbf{s}}_\beta.
\end{align} Thus, the inclination and the surface cannot be uniquely identified. To show the invariance of the column space with inclination, we apply the factorization of the forward model:
\begin{align}
\operatorname{col}(A(\beta)) = \operatorname{col}(W_\omega B_\beta) = W_\omega \operatorname{col}(B_\beta).
\end{align}
Thus the column space of $A(\beta)$ only varies with the column space of $B_\beta$. As described, $B_\beta$ is  not full rank since $4$ rows are linearly dependent between the sub-matrices $B_\beta^x$ and $B_\beta^y$. The rows corresponding to spherical harmonic order $m = \pm L$ and $m = \pm (L-1)$ produce identical measurements up to a scaling, respectively in $x$ and $y$ measurement directions, reducing $\operatorname{rank}(B_\beta)$. This is because only the $l=L$ spherical harmonics contribute towards the non-zero elements of these rows. To show the invariance of $\operatorname{col}(B_\beta)$, we consider a reduced version of the matrix $B_\beta$ defined as $\bar{B}_\beta = [\bar{B}_\beta^x, B_\beta^y]^T$ where $\bar{B}_\beta^x$ is absent the redundant first and last two rows of $B_\beta^x$. This new matrix $\bar{B}_\beta \in \mathcal{C}^{(4L-2) \times (L+1)^2}$ has full row rank of $4L-2$. Since $\bar{B}_\beta$ has full row rank $4L-2$ for all $\beta \in (0,\pi/2)$, its column space is the entire space $\mathcal{C}^{4L-2}$, independent of $\beta$. The reduced matrix $\bar{B}_\beta$ has the same information content as the full matrix $B_\beta$. Because the column space of the reduced matrix is invariant with $\beta$, so too is the column space of the full $B_\beta$ and therefore $A(\beta)$. Thus, any inclination $\beta$ can produce an astrometric signal $\mathbf{\mu}$ and so cannot be uniquely identified. 

To jointly estimate the pair $(\mathbf{s}, \beta)$, it is necessary to introduce a prior constraint on the surface solutions $\mathbf{s}$ that breaks the degeneracy among surface solutions. This ensures the uncertainty of the inclination $\beta$ is bounded \citep{taaki_icassp}. We do not additionally place a prior on the inclination; however, if partial information were known (such as from a $v \sin i$ measurement \citep{vsini}), this could be introduced as an additional constraint. The use of a prior might seem to imply an artificial bias of the surface solutions, but the properties of the prior needed to break the degeneracy can be very weak in practice. For example, a constraint that prevents a single spherical harmonic mode from dominating the solution will effectively steer the solution away from unphysical surfaces. The optimal choice of prior, which brings us closest to the true stellar inclination, will depend on the physical properties of the surface. In Section \ref{sec: crb}, we discuss the choice of surface prior applied herein.

\section{ASTROPHYSICAL IMPLICATIONS} \label{sec: results}
Above, we described the forward model for astrometric jitter measurements of a rotating star. Section \ref{sec: jitterlevel} explores the implications of astrometric jitter by comparing the expected astrometric jitter amplitudes for main-sequence stars to the reflex motion amplitudes of habitable-zone exoplanets. In Section \ref{sec: crb} we describe our approach to invert astrometric measurements and map the stellar surface. In Section \ref{sec: recon}, we demonstrate stellar surface mapping for astrometry and photometry, individually and combined. 

\subsection{Starspot jitter and exoplanet astrometry} \label{sec: jitterlevel}
Astrometric jitter is expected to be a limiting factor in measuring the masses of Earth analogs orbiting in the habitable zone (HZ). Here we introduce a simple framework for comparing astrometric jitter as a function of starspot size to planet-induced reflex motion, in order to determine when jitter dominates. Furthermore, we use the astrometric jitter scaling with spot size to bound regions of the parameter space where starspots can be mapped with Gaia DR4 and DR5 precision. 


From \cite{shao2009astrometric}, $\alpha$ is the amplitude of the astrometric reflex motion of a star with mass $M_\star$ at a distance $D$ with a planet $M_p$ orbiting with semi-major axis $a$:
\begin{align} \label{eq: astr_amp}
\alpha = 3 \frac{M_\odot}{M_\star} \frac{M_p}{M_\oplus} \frac{a}{1\,\text{AU}} \frac{1\,\text{pc}}{D} \, \mu \text{as}.
\end{align}
For each main-sequence radius $R_\star$, we place the representative habitable-zone (HZ) semi-major axis at $a_{\rm HZ}\propto \sqrt{L_\star / L_{\odot}}\propto (M_\star / M_\odot)^{1.75}$, defined as the distance at which the incident stellar flux equals that received by Earth from the Sun \citep{direct_book}. Evaluating the astrometric amplitude expression in Equation \ref{eq: astr_amp} with $a_{\rm HZ}$ and using the mass-radius relation, for an Earth-mass planet $M_p = M_{\oplus}$:
\begin{align} \label{eq:alpha_hz}
\alpha_{\rm HZ} = 3\frac{1\,\text{pc}}{D} \left( \frac{R_{\star}}{R_\odot} \right)^{0.94} \mu \text{as}.
\end{align}
To detect an Earth-mass planet orbiting in the HZ of a solar-mass star at 10 pc would require astrometric precision of $\alpha_{\rm HZ} = 0.3 \mu \text{as}$. 

The astrometric jitter amplitude for a single starspot follows the model introduced by \citet{morris2018spotting}. We assume a single starspot at an equatorial latitude, the fractional flux change due to a starspot of radius $R_{\rm spot}$ is approximately $F_{\rm spot} / F_\star = C R_{\rm spot}^2 / R_\star^2$ where $C$ is the wavelength-dependent spot-contrast. Starspot contrast is largest at bluer wavelengths \citep{sag}. We model a Gaia $G$-band ($330$–$1050$\,nm) and adopt a G-band spot contrast of $C=0.7$ based on \citet{morris2018spotting}. Photo-center shift additionally scales with the angular radius of the star $R_\star / D$, the maximum astrometric jitter due to a starspot is $\alpha_{\rm spot} \approx \frac{F_{\rm spot}}{F_\star} \frac{R_\star}{D}  4650 \, \mu \text{as}$. This jitter model becomes inaccurate for spots that are large compared to the radius of the star. We consider starspots with $R_{spot}$ up to $0.3 R_\star$ for which the astrometric approximation error is $< 6\%$ \citep{morris2018spotting}.

\begin{figure*}[ht!]
\centering
\includegraphics[width=.7\linewidth]{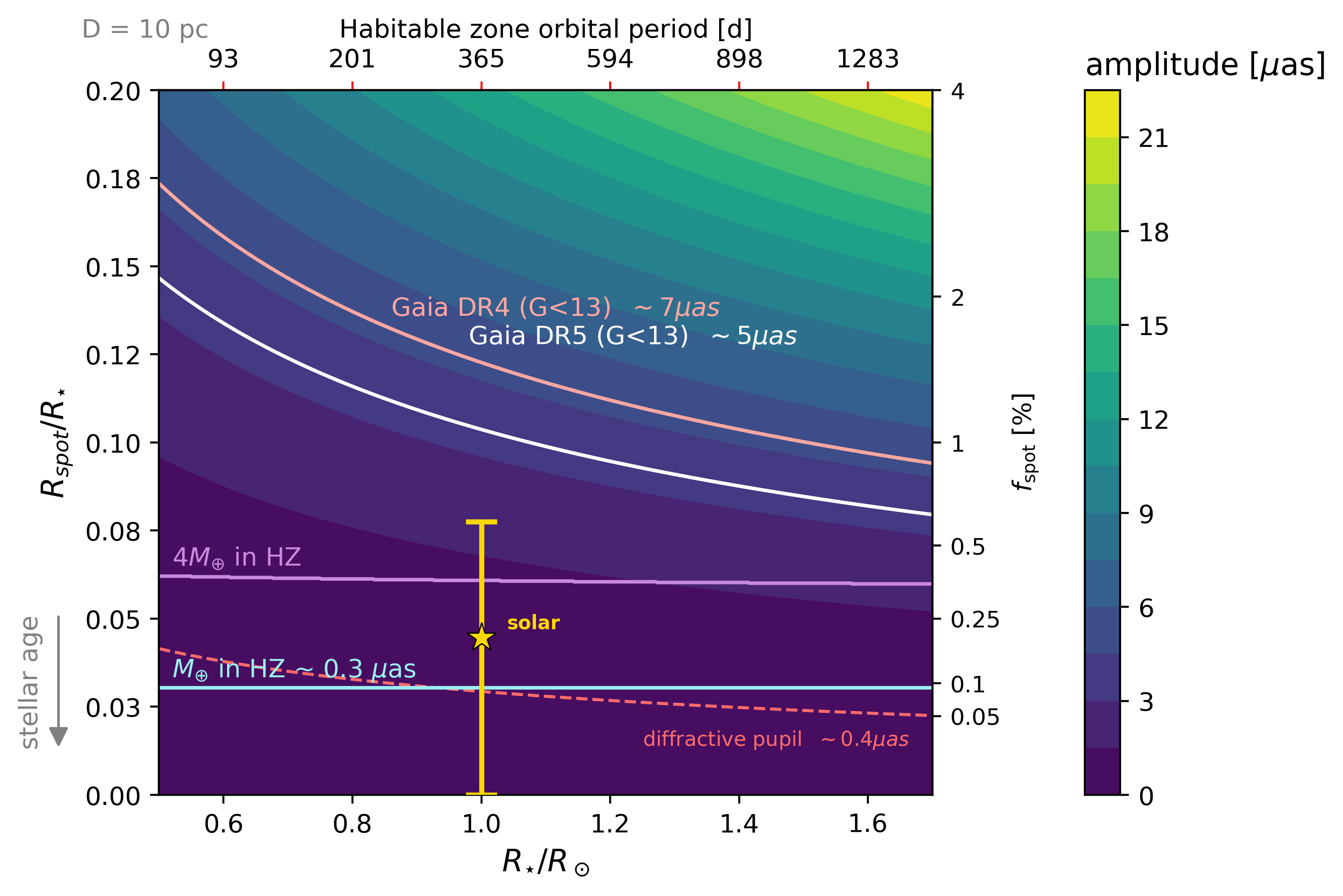}
\caption{The jitter amplitude for a single starspot is shown as a heatmap as a function of starspot radius (y-axis) and stellar radius (x-axis), for a star at a distance of $D=10$ pc. The right y-axis shows visible disk coverage $f_{spot} = (R_{spot} / R_\star )^2$. The Gaia DR5 precision over the G-band for a $G=13$ magnitude star is also overlaid (dark orange), showing that if large starspots occur on these stars, they can be detected. For reference, the effective solar starspot coverage is shown between its activity minimum and maximum. Younger stars, as well as less massive stars, are more magnetically active and are therefore expected to have higher levels of starspot coverage. The astrometric precision for a subset of bright, nearby Gaia stars that are candidates for mapping, from \citet{morris2018spotting}, is overlaid (white). For these stars, with starspots whose radii lie above this line, these are large enough to be mapped with Gaia data.Astrometric jitter may additionally prohibit precise mass measurements of Earth analogs orbiting in the habitable zone (HZ). While Gaia is not sensitive enough to measure small HZ planets, future instruments aim to accomplish this. We illustrate this limiting error; overlaid are HZ exoplanet signal amplitudes for sub-Neptune-sized planets. The top x-axis shows the orbital period of a planet in the HZ as it scales with stellar radius. For starspots where the spot radius is above the exoplanet amplitude lines, astrometric jitter is larger in magnitude than the planet signal.  \label{fig: gaia}}
\end{figure*}

Figure \ref{fig: gaia} illustrates the potential range in starspot-induced astrometric jitter as a heatmap where we span F,G,K,M stellar radii (0.5 - 1.7 $R_\odot$) over the x-axis and starspot radius on the y-axis. To compare astrometric jitter to exoplanet signals, reflex motion amplitudes, $\alpha_{HZ}$ are overlaid as lines on the heatmap. A starspot larger than 3\% stellar radius will produce astrometric jitter of greater amplitude than an Earth-mass exoplanet orbiting in the habitable zone. This threshold does not change with stellar distance $D$, as both the astrometric jitter and the reflex motion amplitude are inversely proportional to $D$. A single starspot of size $3 \% R_\star$ will cover $0.1 \%$ of the visible hemisphere. Thus, at the Earth-mass level, starspot jitter can degrade mass measurement precision. 
In blind detection tests for a sample of 55 F,G,K stars, an average mass measurement uncertainty of 30\% was found at the detection threshold \citep{meunier}. Mass is estimated from the amplitude of planetary reflex motion, while detecting a planet relies on identifying a periodic reflex-motion signal. Since the planetary reflex motion is well-separated in period from the stellar rotation, detection is robust to astrometric jitter, but jitter still poses challenges for mass measurement \citep{sun-jitter, lagrange2011, meunier2020, meunier}.

Overlaid on Figure \ref{fig: gaia} is the solar starspot coverage level for comparison. Solar starspot coverage is the result of multiple small starspots, whose astrometric jitter contributions may partially cancel and produce lower-amplitude jitter than a single large starspot of equivalent coverage. Thus, the single starspot model shown here represents an upper bound on the astrometric jitter amplitude for a given coverage level and is intended to provide a heuristic amplitude scaling. More detailed studies of solar astrometric jitter, which account for realistic distributions of multiple active regions, have been performed by \citet{sun-jitter, Sowmya_2021, Sowmya_2022,lagrange2011}. The maximum coverage is computed from monthly averages of the total solar starpot area over the full surface \citet{Hathaway_2015} \footnote{\url{https://solarscience.msfc.nasa.gov/greenwch/sunspot_area.txt}}, giving a visible hemisphere coverage between 0.02$\%$-0.6$\%$. 

Astrometric jitter is unlikely to be a significant source of uncertainty for the planet populations observable with Gaia since it is largely insensitive to low-mass exoplanets. As given by Equation \ref{eq: astr_amp}, reflex motion amplitude decreases with planet mass. Considering low-mass planets that produce weak astrometric signals, a Neptune-mass planet orbiting at 4 AU around a solar-mass star will need to be within $D = 3.5$ pc to have a greater than 50\% likelihood of detection in DR5 with $\alpha \sim 50 \mu$as \citep{exop_gaia}. While detecting these low-mass planets in Gaia data is likely to be rare, more massive exoplanets will produce signals far stronger than reasonable jitter amplitudes.  



Figure \ref{fig: gaia} further serves to illustrate the level of starspot coverage measurable by Gaia. \citep{morris2018spotting} show that only nearby ($<$10 pc) active M/K dwarfs, with high levels of starspot coverage, would have detectable star-spot induced astrometric jitter in forthcoming Gaia data release (DR4). We plot the end-of-mission Gaia DR4 and DR5 precision as curves in Figure \ref{fig: gaia}. Large enough starspots that lie above the curve correspond to measurable jitter and thus a mappable surface. While the Sun's total starspot coverage is low $f_{\rm spot} \lesssim 1\% $ \citep{Hathaway_2015} and would be below Gaia's measurement precision, larger regions of coverage have been commonly observed for magnetically active stars \citep{strass}. For example, \citet{Araujo_2025} find starspot coverage between 4-29$ \%$ in a sample of 11 F,G,K,M stars.  Active nearby M dwarfs are a promising population for applying the astrometric mapping technique to probe magnetic activity. In addition, models of Solar-like stars predict that active G dwarfs with rotation periods of a few days can exhibit signatures of around $15$ $\mu$as at 10 pc, detectable with Gaia \citep{Sowmya_2022}. Within 10 pc, Gaia contains 285 F,G,K,M stars, 74 \% of which are are M-dwarfs. In addition to astrometry data, Gaia provides simultaneous photometry and for nearly all of this population (98 \%) TESS photometry is additionally available. Red supergiants (not shown over the range in Figure \ref{fig: gaia}) may show significant astrometric variability in Gaia data \citep{redgiant}. Astrometric variability from stellar surface structures has been reconstructed for the giant XX Trianguli of up to 24 $\mu$as \citep{xxtrianguli}. At even larger signal amplitudes, convection cells some 30 mas in size have been detected on Betelgeuse via interferometry \citep{betel}.

\subsection{Surface inversion} \label{sec: crb}
Starting with the signal model in Equation \ref{eq: main}, we describe the procedure for estimating a surface brightness map of the star from astrometric and photometric data. We first obtain a joint-likelihood function for the data, starting with the signal model and incorporating measurement noise, as well as a prior on the surface components to break the inclination degeneracy described in Section \ref{sec: nullspace}. Assuming additive, independent Gaussian noise with per-epoch variance $\sigma^2$, $\mathbf{n}\sim\mathcal{N}(\mathbf{0},\sigma^2 I)$, the astrometric measurements are given by:
\begin{align}  \label{eq: model}
\mathbf{y} = A(\beta) \mathbf{s} + \mathbf{n}.
\end{align}
We assume the rotation period $P$ of the star, or equivalently the rotation rate $\omega = \frac{2 \pi}{P}$, is known or can reasonably be estimated via spectral techniques applied to light-curve or astrometry data \citep{rotrate, rot_kep}. We use a Gaussian-Markov random field (GMRF) prior on the surface coefficients of the form \(\mathbf{s} \sim \mathcal{CN}(0, \Sigma)\), where \(\Sigma\) is a diagonal covariance matrix that parameterizes a Gaussian–Markov random field. Several features of this prior make it suitable for describing a stellar surface. First, the prior is isotropic and does not bias the inversion towards a particular surface orientation. For a prior to be isotropic, it should be invariant at any location on the sphere. A Gaussian prior is completely parameterized by its mean and covariance. Thus, we can show it is isotropic by inspecting the rotation invariance of these parameters. The mean $\boldsymbol{\mu}_s$ (here taken to be 0) should be independent of spatial position and therefore constant. The covariance $\Sigma$ in spherical harmonic coordinates must be invariant under any rotation. Since rotations are separable over degree $l$, we obtain a rotation of degree $l$ coefficients $\mathbf{s}_l$ through the Wigner D rotation matrix: $D^l(R) \mathbf{s}_l$. This is a linear transform of a Gaussian random vector, and therefore, the distribution of the rotated random vector is also Gaussian. Considering the covariance restricted to the degree $l$ terms, the covariance of the rotated surface is $D^l(R)\Sigma_l D^{l, T}(R)$, and for this to be invariant under rotation, it must equal $\Sigma_l$. Only a scaled identity matrix of the form $\Sigma_l = c_l I$ for some constant $c_l$ will satisfy this rotation invariance.  Equivalently, the spherical harmonic coefficients must be uncorrelated \citep{ieee_sph}.

We next motivate the specific choice of degree-dependent scaling $c_l$.
The GMRF prior corresponds to a quadratic penalty on the gradient of the surface in spatial coordinates, encouraging locally correlated structure while suppressing small-scale fluctuations. Under the GMRF prior, degree-dependent weights penalize higher degree coefficients. This is appropriate, since the astrometric kernel weights decay with higher degree terms that correspond to finer spatial scales, and these components have lower SNR. Thus, downweighting these reduces the solution's sensitivity to measurement noise.  In particular, the GMRF parameterization used here per degree is $\Sigma_l \propto (1/ \lambda)(\gamma/l)^{\kappa}$, $\gamma$ is an angular scale parameter, for $\kappa > 1$. $\gamma$ sets the characteristic degree at which regularization transitions: harmonics below degree $\gamma$ have larger prior variance and are less penalized, while those above $\gamma$ are progressively suppressed; $\kappa$ controls how steeply the suppression increases with degree, corresponding to an angular scale of ${\sim}180^\circ/\gamma$. Here, $\lambda > 0$ is a regularization weight that controls the relative strength of the prior versus the data likelihood term. A similar scale-dependent prior is used in \citet{farr} in the context of mapping exoplanets in reflected light, to enforce spatial correlations over spherical harmonic coordinates.

To obtain an estimate for the unknown surface and inclination, we maximize the likelihood of Equation \ref{eq: model}, which is equivalent to minimizing the negative log-likelihood given by:
\begin{equation}\label{eq: likelihood}
-\log p(\mathbf{y},\mathbf{s}\mid \beta)\ \propto\
\underbrace{\frac{1}{\sigma^2}\,\big\|\mathbf{y} -A(\beta)\mathbf{s}\big\|_2^2}_{\text{data penalty}}
\;+\;
\underbrace{\mathbf{s}^{H}\Sigma^{-1} \mathbf{s}}_{\text{GMRF penalty}}.
\end{equation}
This likelihood function consists of a data penalty term that depends on the data $\mathbf{y}$, and a GMRF penalty term on $\mathbf{s}$. To jointly minimize the negative log-likelihood over the surface coefficients $\mathbf{s}$ and inclination $\beta$, we treat $\beta$ as fixed, evaluating it over a discrete grid of values in $[0, \pi/2]$, and solve for the optimal surface $\hat{\mathbf{s}}_\beta$ at each. We then select the pair $(\beta, \hat{\mathbf{s}}_\beta)$ that minimizes the negative log-likelihood from among these solutions. For a fixed inclination $\beta$, the negative log-likelihood is minimized over $\mathbf{s}$ by the regularized pseudoinverse of \( A(\beta) \) applied to the data $\mathbf{y}$:
\[
\hat{\mathbf{s}}_\beta = \left( A(\beta)^H A(\beta) +  \Sigma^{-1} \right)^{-1} A(\beta)^H \mathbf{y}.
\]
These surface estimates $\hat{\mathbf{s}}_\beta$ will not have any component in the null space. This is because the GMRF penalty term in Equation \ref{eq: likelihood} is non-negative for any $\mathbf{s}$ and will increase for a null-space component, while having no effect on the data penalty. Therefore, a solution with a component in the null space will not be a minimizer of the negative log likelihood. We note the mean-square error between the estimated and true coefficients, MSE$ = \|\hat{\mathbf{s}} - \mathbf{s}\|_2^2 = \|\sum_{l,m} Y_l^m (\hat{s}_{l}^m - s_{l}^m) \|_2^2$ is equivalent to the surface estimation error, since the spherical harmonic transform is orthogonal and therefore preserves norms on vectors. 

\subsubsection{Combining astrometry and photometry} \label{sec: combined}
As noted in Section~\ref{sec: kernel}, the photometric forward model has the same time and inclination separable structure as the astrometric model (Equation~\ref{eq: main}):
\begin{align} \label{eq: phot_model}
\boldsymbol{\mu}^{\mathrm{phot}} = W_\omega\, B_\beta^{\mathrm{phot}}\, \mathbf{s},
\end{align}
where $B_\beta^{\mathrm{phot}}$ is constructed from the photometric kernel $\mathbf{k}^{\mathrm{phot}}$ (Appendix~\ref{ap: photo}), the zeroth-moment analogue of the astrometric kernel. We note that the time-dependent matrix $W_\omega$ can be specified for different observation times between photometric and astrometric measurements.

To combine astrometric and photometric measurements, we stack the individual forward models into a single measurement equation, however allowing for different noise levels. Let $\mathbf{y}_a = [\mathbf{y}^x\, \mathbf{y}^y]^T$ denote the astrometric data with per-epoch noise variance $\sigma_a^2$, and $\mathbf{y}_p$ the photometric data with noise variance $\sigma_p^2$. Labelling $A_{\mathrm{phot}}(\beta) = W_\omega B_\beta^{\mathrm{phot}}$ and $A_{\mathrm{astr}}(\beta)$ is defined in Equation~\ref{eq:model}, the joint measurement model is:
\begin{align} \label{eq: joint_model}
\underbrace{
\begin{bmatrix}
\mathbf{y}_a \\[2pt] \mathbf{y}_p
\end{bmatrix}}_{\mathbf{y}}
\;=\;
\underbrace{
\begin{bmatrix}
A_{\mathrm{astr}}(\beta)\\[2pt] A_{\mathrm{phot}}(\beta)
\end{bmatrix}}_{A(\beta)}
\mathbf{s}
\;+\;
\underbrace{
\begin{bmatrix}
\mathbf{n}_a \\[2pt] \mathbf{n}_p
\end{bmatrix}}_{\mathbf{n}},
\end{align}
Defining $\tilde{A}(\beta) = [A_{\mathrm{astr}} / \sigma_{a};\; A_{\mathrm{phot}}/\sigma_p]$ and $\tilde{\mathbf{y}} = [\mathbf{y}_a / \sigma_a;\; \mathbf{y}_p / \sigma_p]$, the minimizer of the joint log likelihood for both measurement types and a fixed value of $\beta$ is:
\begin{equation}\label{eq: joint_map}
\hat{\mathbf{s}}_\beta = \left(\tilde{A}(\beta)^H \tilde{A}(\beta) + \Sigma^{-1}\right)^{-1}\tilde{A}(\beta)^H \tilde{\mathbf{y}}\,.
\end{equation}
The noise variances $\sigma_a^2$ and $\sigma_p^2$ control the relative influence of each data type on the estimates of the $l=1$ and $l=2$ spherical harmonics that are in the shared measurement space. For the remaining spherical harmonics, because astrometry and photometry measure different sets of spherical harmonics, the estimates are independent between the two data types. Thus, the relative noise level does not affect the estimate, and the individual specified noise variances affect the estimate only in the sense that they control the tradeoff between data and prior. In the reconstruction examples that follow (Section~\ref{sec: recon}), we demonstrate this joint inversion as a proof of concept with equal noise levels across channels.

\subsection{Reconstruction examples} \label{sec: recon}
\begin{figure}
\centering
\includegraphics[width=\textwidth]{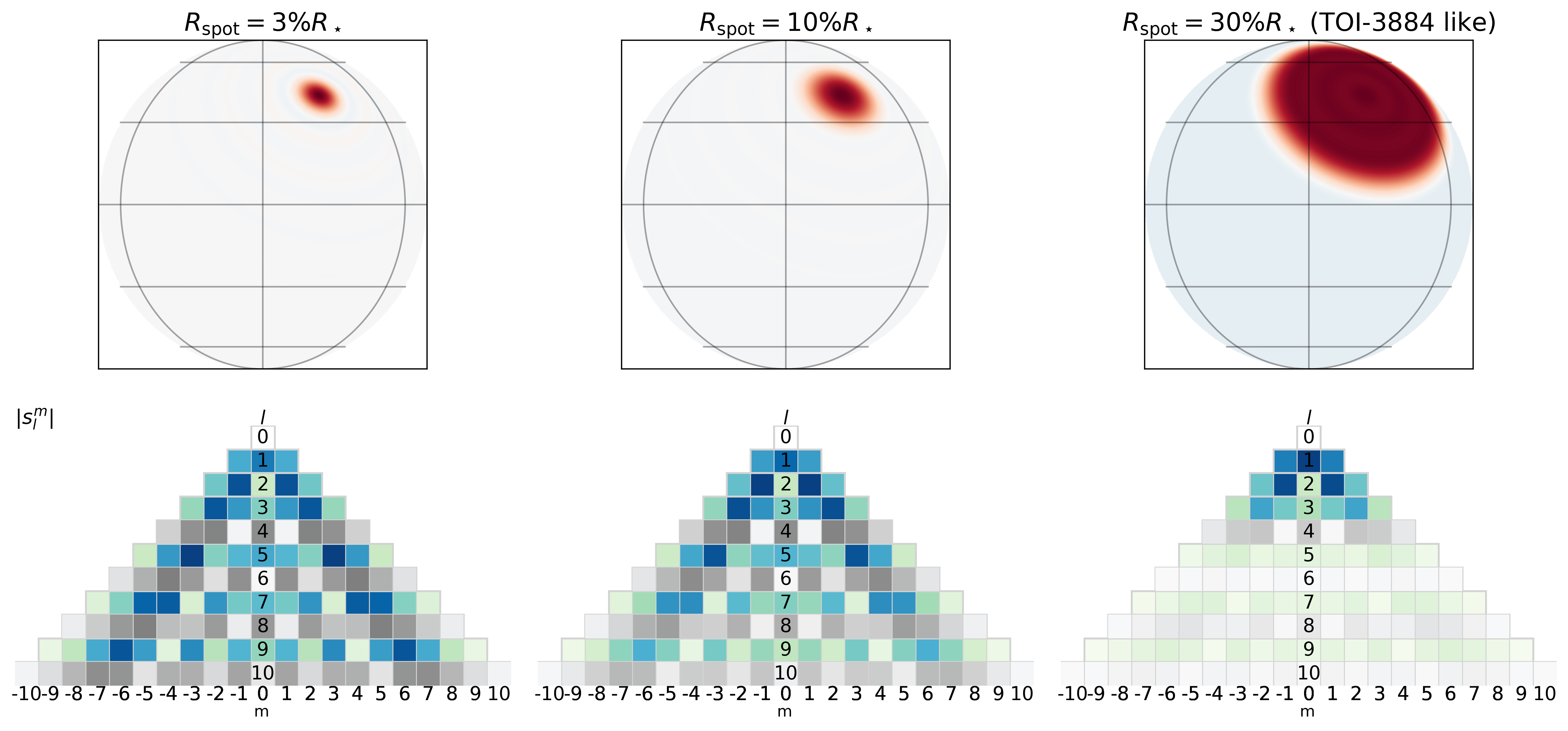}
 \caption{Visualizing the starspot coefficients for starspots of varying radii. The surface is shown equator-on up to degree $L = 25$ (top row) and the corresponding surface coefficients $\mathbf{s}$ up to degree $L=10$ (bottom row). The surface coefficients that do not contribute to the astrometric measurements and belong to the nullspace are shown in grey. Additionally, we have set the mean flux term to zero ($s_0^0 = 0$), as it is much larger in magnitude than the other signal components. These illustrate the Fourier uncertainty principle: smaller, more compact starspots have larger spherical harmonic representations. } \label{fig: starspot_coeff}
\end{figure}

\begin{figure*}
\nolinenumbers
\figsetstart
\figsetnum{9}
\figsettitle{Reconstruction gallery}
\figsetend
\linenumbers

\centering
\includegraphics[width=1\textwidth]{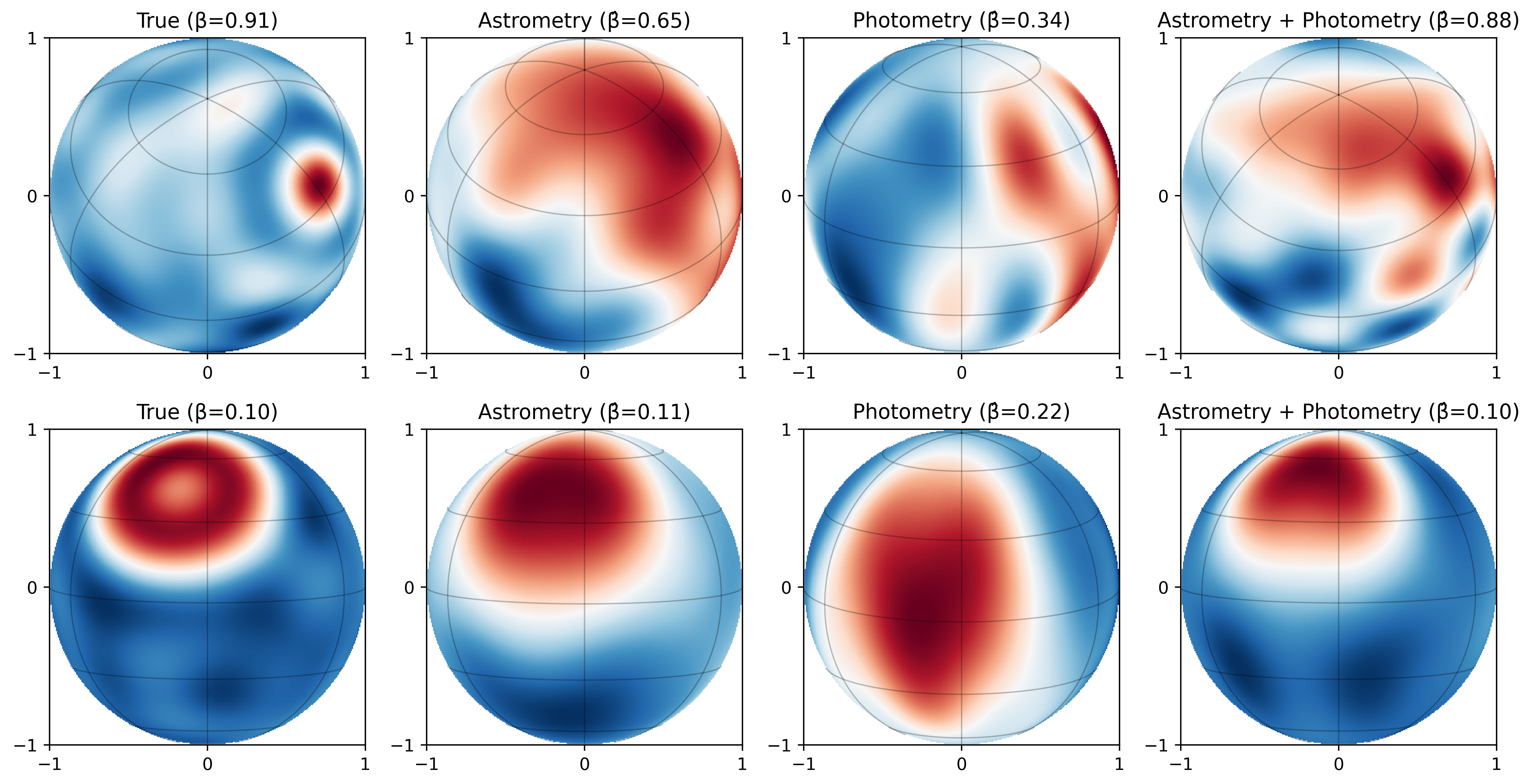}
 \caption{Two example reconstructions for a simulated $8 \%$ $R_\star$ starspot (top row) and a TOI-3884 like $30 \% R_\star$ starspot (bottom row). The first column shows the ground truth simulated surface at the true inclination angle, represented by a truncated spherical harmonic expansion up to $L = 10$. Reconstructions of the surface and inclination $\beta$ (assumed unknown) are shown using astrometry mapping (column two), photometry mapping (column three), and combined astrometry-photometry mapping (column four). All solutions use the same regularized least-squares inversion approach described in Section \ref{sec: crb}. Here, the photometric surface reconstructions tend to be smeared latitudinally, reflecting an inherent degeneracy in identifying spot latitude and stellar inclination with photometry. The astrometric reconstruction shows a localized starspot and better constraints on stellar inclination. Combining astrometry and photometry improves the accuracy of inclination and latitude by providing complementary information. A supplementary gallery showing reconstructions across a variety of spot configurations and signal-to-noise ratios is available in the online journal. } \label{fig: recon}
\end{figure*}

In this section, we perform example surface inversions with simulated surfaces, astrometry and photometry data. We present two representative reconstructions in the main text, alongside a supplementary gallery spanning spot sizes of $5-25\% R_\star$, inclinations and multiple SNR levels, available as an online figure set (Figure \ref{fig: recon}). For the reconstructions shown here, we simulate surfaces with a single starspot, generated with Starry \citep{Luger_2019}, with the mean flux set to zero by the $l=m=0$ spherical coefficient. In Figure \ref{fig: starspot_coeff}, the spherical harmonic coefficients for starspots of varying radii are visualized. Larger starspots have compact spherical harmonic representations, whereas smaller starspots have many non-zero coefficients. This behaviour is also observed in Fourier transforms  \citep{fourier_space}. We simulate the astrometric and photometric signals for two surfaces with different starspot latitudes, sizes, and stellar inclinations. Each surface contains a single starspot, and to each we add a small randomly generated GMRF component to emulate variation from smaller active regions. The first surface has an $8 \% R_\star$ starspot at a latitude of $35^\circ$. The second has a larger $30 \% R_\star$ starspot at a latitude of $50^\circ$, motivated by TOI-3884, an M-dwarf at 43 pc with a large near-polar starspot that was detected through repeated transit crossings of TOI-3884b in TESS light curves \citep{toi3884, tamburo}. Transit crossings of starspots can help constrain their latitudinal positions \citep{Sagynbayeva_2025}.
The astrometric jitter is of the order $\alpha \sim 0.3 \mu$as, below Gaia's sensitivity, as the star is at a distance of 43.34 pc and the spot is at a near-polar latitude.

When performing the surface inversion, the maximum spherical harmonic degree $L$ must be specified. The number of observations $N$ both limits the signal-to-noise ratio, as well as the maximum degree $L$ that can be resolved without aliasing. According to the Shannon-Nyquist sampling theorem \citep{oppenheim1993digital}, to resolve a signal without aliasing, the sampling frequency $f_s$ must satisfy $f_s > 2 f$, where $f$ is the highest frequency component of the signal. The astrometric time-series signals in $x$ and $y$ (Equation \ref{eq: main}) are sinusoids composed of conjugate sums of complex exponential oscillations. The highest temporal frequency present occurs for a spherical harmonic of order $m = L$, to resolve this, we then require $N > 2 L$. Equivalently, if we have some $N = 31$ measurements, we may only resolve up to $L = 15$ spherical harmonics. If $L > \frac{N}{2}$, the contribution of higher-order harmonics will appear as if a lower-order term created it, so smaller-scale surface features will project to larger-scale terms. In this example, we simulate $N = 100$, comparable to the number of along-scan epochs per star expected in Gaia DR5, and choose $L = 10$ for both the simulation and inversion, below the aliasing cutoff. Without loss of generality, we take the orbital period as $P = 1$.

For each simulated surface, we apply the Bayesian approach described in \ref{sec: crb} to estimate the stellar surface and inclination, using the same GMRF prior and hyperparameter values for both photometry and astrometry data. We use the following parameter values: $\lambda = 10^{-4}, \gamma = 2, \alpha = 1$. The GMRF prior used here is invariant with spatial position and can encompass a broad range of surface features. We use this inversion technique for astrometry alone, for photometry alone, and for combined astrometry and photometry data.
To assess the impact of measurement noise in the figure set, for each simulated signal we add i.i.d. Gaussian noise (Equation \ref{eq: model}). Defining the signal-to-noise ratio as $\mathrm{SNR} = \| \boldsymbol{\mu}^h \|_2 / \sigma$, where $\|\boldsymbol{\mu}^h \|_2$ is the amplitude of the noiseless signal and $\sigma$ is the per-epoch noise standard deviation, we present reconstructions at three SNR. These are noiseless, $\mathrm{SNR} = 100$, and $\mathrm{SNR} = 10$. These are chosen to based on the current and future instrument precisions in Section \ref{sec: jitterlevel}, $\mathrm{SNR} \lesssim 10$ is consistent with the Gaia end-of-mission precision for a nearby ($D < 10$ pc) star with a large starspot ($\geq 25\% R_\star$) \citep{morris2018spotting}, while $\mathrm{SNR} \sim 100$ is achievable with future sub-$\mu$as astrometry for moderate spot sizes ($\geq 10\% R_\star$). We apply the same SNR to both the astrometric and photometric channels for comparison. In practice, for PSF based astrometry such as Gaia, or with data from Kepler/TESS, the photometric precision is significantly better than the astrometric precision.

The ground truth simulated stellar surfaces and noiseless reconstructions are shown in Figure \ref{fig: recon}. For these example figures, we show only the noiseless case, representing the theoretical reconstruction limit.  Identifying spot latitude and stellar inclination with photometry is known to be degenerate. In the reconstructions from photometry, the surface features are smeared latitudinally, reflecting this uncertainty. Astrometry reconstructions provide better constraints on stellar inclination and spot latitude than photometry reconstructions. When photometry and astrometry data are combined, inclination accuracy improves further, as does the overall surface reconstruction accuracy.

\section{CONCLUSION} \label{sec: discussion} \label{sec: conclusion}

Although astrometric jitter is a source of noise in measuring exoplanet-induced reflex motion, we show that jitter can be inverted to map stellar surfaces. 
To show this, we derived an analytic forward model for astrometric jitter induced by starspots on an inclined, rotating stellar surface. The model is linear in the spherical-harmonic representation of the stellar surface (Section \ref{sec: methods}). The astrometric response to each static spherical harmonic is the astrometric kernel, which serves as the foundation of this model. A full astrometric time series is generated as a linear combination of the astrometric kernel terms. 

Light-curve inversion has been widely used for mapping stellar surfaces, but it suffers from degeneracies, particularly in constraining spot latitude \citep{ekeri, kovari, Walkowicz_2013, Davenport_2015, rottenbacher}. Using the astrometric kernel, we demonstrate that astrometry (first moments of a stellar point-spread function) measures a different set of spherical-harmonic surface modes than light curves (zeroth moments), thereby breaking degeneracies in surface mapping with photometry alone. Light curve data measures even-degree $l$ harmonics that are symmetric about the equator \citep{cowan}. In contrast, astrometry measures odd-degree $l$ harmonics that are antisymmetric about the equator and introduce asymmetries in the photo-center. Furthermore, astrometry measures two independent components of the photo-centre motion (x and y), constraining both latitude and longitude of a starspot. In spherical-harmonic terms, for a star viewed from the equator, among odd-degree $l$ harmonics, the $x$ direction component measures odd-order $m$ harmonics and the $y$ direction, even-order $m$ harmonics, while photometry only measures even degree $l$ and even-order $m$ harmonics. 

For a star with an inclined rotational axis, photometry and astrometry each contribute unique information about the stellar surface by still only observing even- and odd-degree spherical harmonics, respectively. This means that when astrometry and photometry measurements are combined the surface inversion accuracy is improved. However, the sensitivity of photometry and astrometry to a particular spherical harmonic order $m$ changes with inclination. Through quantifying the resulting change in the rank (the number of recoverable degrees of freedom in the forward model), we show that interior inclinations $\beta \in (0, \pi/2)$ carry the most recoverable surface information. A larger rank equates to a smaller nullspace, meaning fewer possible stellar surfaces can produce the measured astrometric signal, and the surface inversion is better constrained. Interior inclinations achieve a substantially higher rank ($\sim 4L$) than equator-on ($\sim 2L$) or pole-on limits ($2$), which carry less information for both astrometry and photometry. Viewing the star pole-on carries the least surface information, and any astrometric signal will be purely circular, while photometry can only obtain the mean flux. 

When the stellar inclination is unknown, it is unidentifiable without any external constraints: for every candidate inclination there exists a surface that exactly reproduces the observed astrometric or photometric signal.  In Section \ref{sec: crb}, we introduce a Bayesian inversion approach to estimate both stellar inclination and the surface brightness map of the star. To render the inclination identifiable, we introduce a spatial-scale-dependent prior on the stellar surface \citep{taaki_icassp}. This Gaussian-Markov-Random-Field (GMRF) prior encourages locally correlated, piecewise-smooth structure while still permitting sharp discontinuities in the surface amplitude, for example, at the edge of a spot. 

We perform example surface inversions with both light curves and astrometry data under the Bayesian GMRF prior. Our simulations show that astrometry retrieves two directions of information and can reliably localize the spot position alongside stellar inclination, whereas photometric reconstructions tend to be smeared latitudinally. When astrometry and photometry are combined, overall reconstruction accuracy improves. Constraints on stellar inclination through astrometric jitter may provide a new angle for studying spin-orbit misaligned systems \citep{Albrecht_2022}. 

The nearby population of M-dwarfs ($<$ 300 stars within 10 pc) accessible to Gaia provides an upcoming opportunity to apply the framework developed here. For stable spot configurations, Kepler and TESS data may offer complementary photometry constraints. Because starspots also bias radial-velocity and transmission-spectroscopy measurements \citep{sag}, improved stellar surface mapping can help correct biases across various measurements for the mapped stars.

As shown in Figure \ref{fig: gaia}, a single starspot with a radius $3\% R_\star $ or larger may create jitter at the amplitude as the reflex motion due to an Earth-mass exoplanet orbiting in the habitable zone. While interior inclinations are optimal for surface mapping, \citet{Sowmya_2021} find that these non-equatorial astrometric jitter signals are harder to disambiguate from planet-induced reflex motion. Non-equatorial inclinations, however, are optimal for direct imaging of exoplanets in reflected light. Future direct-imaging missions, such as the Habitable Worlds Observatory, can take days to weeks to image an exoplanet in reflected light \citep{mennesson}. As \citet{Painter_2025} describes, they are most efficient when they can target systems with precursor astrometric detections of planets and constrained planetary masses. Our model quantifies the variations in astrometric jitter with inclination angle, providing an analytical foundation for studying starspot noise in astrometry, worst-case geometries, and their impact on exoplanet mass measurements in future sub-$\mu$as astrometry.  In future work, we aim to use the inversion framework to jointly fit simulated sub $\mu$as reflex motion and astrometric jitter and to quantify the impact on mass-estimator precision. 
 

A limitation of this work is that we have not incorporated limb darkening or extended our analysis to multiple observing bands. These factors may change the identifiability of surfaces. \citet{Luger_2021} consider quadratic limb darkening in the context of photometry and find a non-trivial impact on the forward model, which varies with the assumed limb-darkening model. In future work, we will incorporate quadratic limb darkening into the astrometric operator and obtain an astrometric kernel under this setting. Besides limb darkening, stars exhibit differential rotation and have non-static surfaces. \citet{Davenport_2015} use four years of Kepler observations to detect differential rotation and starspot evolution. By breaking latitude degeneracies, astrometric surface mapping may offer a valuable probe of differential rotation, but extending our framework to time-variable surfaces will be necessary to interpret longer-baseline observations, as expected with upcoming Gaia data.

\begin{acknowledgments}
This project is supported by the Eric and Wendy Schmidt AI in Science Postdoctoral Fellowship, a program of Schmidt Sciences. This research was partially supported by grant CCF-2246213 from the National Science Foundation.
\end{acknowledgments}
\software{Sympy \citep{symp}, Starry \citep{Luger_2019}}

\appendix
This appendix derives the forward model summarized in Section \ref{sec: methods} and provides the technical details. We first describe the rotation formalism used to express astrometric first moments in spherical-harmonic space, including the decomposition that isolates the inclination angle (Appendix \ref{ap: rot}). We then derive the gradient of the forward model with respect to inclination (Appendix \ref{ap: grad}), followed by the construction of the astrometric kernel and its role in the forward model (Appendix \ref{ap: astr}). 
\section{Inclined rotation} \label{ap: rot}

We define $h \in \{x, y\}$ as the projected x and y-coordinates from the observer's perspective. A first moment measurement $\mu^h : h \in \{x, y\}$ at a rotation $R = (\alpha, \beta, \gamma)$ for stellar surface spherical harmonic coefficients $\mathbf{s}$ is given by:
\begin{align} \label{eq: astr_1}
\mu^h(R) = \sum_{l \geq 0}\ip{\mathbf{s}_l}{D^{l, H}(R) \mathbf{k}_l^h}
\end{align}
The Wigner-D rotation matrix factorizes as:
\begin{align} \label{eq: wigner_def}
 D^l_{mm'}(\alpha, \beta, \gamma) = e^{-i m \alpha}\, d^l_{m, m'}(\beta)\, e^{-i m' \gamma}
\end{align}
where $d^l_{mm'}(\beta)$ is the real-valued small Wigner-d matrix. The physical rotation is $R = (0, \beta, \omega t)$: inclination $\beta$, and spin $\gamma = \omega t$. 

Since $D^l$ is unitary, the conjugate transpose in Equation~\ref{eq: astr_1} equals the inverse rotation, which swaps and negates the Euler angles:

\begin{align} \label{eq: conj_swap}
D^{l,H}(\alpha, \beta, \gamma) = D^l(-\gamma,\, -\beta,\, -\alpha).
\end{align}
The swap $\alpha \leftrightarrow \gamma$ moves the time dependence (originally in $\gamma = \omega t$) into the first Euler angle, where it appears as a simple phase $e^{-im(-\omega t)} = e^{im\omega t}$

\begin{align} \label{eq: rot_sph}
D^{l, H}_{mm'}(0, \beta, \omega t) =  e^{im\omega t} d^l_{mm'}(\beta).
\end{align}
The time-dependent phase $e^{im\omega t}$ can be factored out of the summation over $l$ in Equation~\ref{eq: astr_1}. Furthermore, the transpose of the small Wigner-d matrix negates its argument: $d^l_{m'm}(\beta) = d^l_{mm'}(-\beta)$. Using this identity, the signal at time $t$ is:

\begin{align} \label{eq: mix}
\mu^h(t) = \sum_m e^{im\omega t} \sum_l \bigl[ d^l(-\beta)\, \mathbf{k}^h_l \bigr]_m\, s_{l,m}.
\end{align}
The matrix $d^l(-\beta)$ mixes orders $m \in \{-l, \ldots, l\}$ within each degree $l$. For $N$ observations at times $t_1, \ldots, t_N$, collecting $\boldsymbol{\mu}^h = [\mu^h(t_1), \ldots, \mu^h(t_N)]^T$:

\begin{align} \label{eq: simple}
\boldsymbol{\mu}^h = W_{\omega}\, B_{\beta}^h\, \mathbf{s}.
\end{align}
The forward model separates into a time-dependent component $W_\omega \in \mathbb{C}^{N \times (2L+1)}$ and an inclination- and observational direction $h \in \{x, y\}$ dependent matrix $B_\beta^h \in \mathbb{C}^{(2L+1)\times (L+1)^2}$. The matrix $B_\beta^h$ encodes the inclination-rotated kernel as a row-block matrix of diagonalized vectors over degree $l$:

\begin{align}
B_\beta^h = \left[ \operatorname{diag}(d^0(-\beta)\, \mathbf{k}_{0}^h),\; \ldots,\; \operatorname{diag}(d^L(-\beta)\, \mathbf{k}_{L}^h) \right].
\end{align}
For a different measurement kernel, i.e., the zeroth moment (photometry), the corresponding kernel can be dropped in to this forward model without any changes.
The matrix $W_{\omega} \in \mathbb{C}^{N \times (2L+1)}$ encapsulates the time dependence. We assume a full rotation is observed with uniform temporal sampling although this can easily be modified to handle arbitrary points in time. $W_\omega $ is a Vandermonde-like matrix whose columns are harmonics:
\[
W_{\omega} =
\begin{bmatrix}
\lambda^{0 \cdot (-L)} & \lambda^{0 \cdot (-L+1)} & \cdots & \lambda^{0 \cdot L} \\
\lambda^{1 \cdot (-L)} & \lambda^{1 \cdot (-L+1)} & \cdots & \lambda^{1 \cdot L} \\
\vdots & \vdots & \ddots & \vdots \\
\lambda^{(N-1) \cdot (-L)} & \lambda^{(N-1) \cdot (-L+1)} & \cdots & \lambda^{(N-1) \cdot L}
\end{bmatrix},
\]
where $\lambda = e^{i 2\pi \omega / N}$. When $N\geq(2L+1)$ this matrix is unitary and $W_\omega^H W_\omega = N I_{(2L+1)}$.

\subsection{Inclination gradient} \label{ap: grad}
We state the analytic gradient of $\boldsymbol{\mu} = A(\beta) \mathbf{s}$ with respect to inclination $\beta$. This gradient can be used for gradient-based optimization of $\beta$ when jointly estimating the surface and inclination (Section~\ref{sec: crb}), and to derive the Fisher information matrix for $(\mathbf{s}, \beta)$, which quantifies the theoretical precision bounds on both estimates \citep{taaki_icassp}.

To compute $\partial B_\beta^h / \partial \beta$, we require the derivative of $d^l(-\beta)$ with respect to $\beta$. Differentiating $d^l_{mm'}(-\beta)$ directly requires recurrence relations. Instead, following \citet{shift_paper}, we decompose

\begin{align} \label{eq: D1D2}
d^l(-\beta) = \bigl[D_1^l \cdot D_2^l(\beta)\bigr]^T,
\end{align}
where $D_1^l = D^l(\pi/2,\, \pi/2,\, 0)$ is a fixed matrix independent of $\beta$, and $D_2^l(\beta) = D^l(\beta+\pi,\, \pi/2,\, \pi/2)$ contains $\beta$ only in its first angle. Since $\beta$ enters $D_2^l$ through the phase $e^{-im(\beta+\pi)}$, the derivative is diagonal:

\begin{align} \label{eq: dD2}
\frac{\partial D_2^l}{\partial \beta} = -\,i\, \operatorname{diag}(-l,\ldots,l)\; D_2^l.
\end{align}
This gives the inclination gradient a simple closed form.

With \(A(\beta)=W_\omega B_\beta\):

\begin{align}
\frac{\partial\boldsymbol{\mu}^h}{\partial \beta} = W_\omega \frac{d B^h_\beta}{d\beta}\,\mathbf{s},\\
\frac{d B_\beta^h}{d\beta}
  = \big[\operatorname{diag}(\tfrac{d}{d\beta}[d^0(-\beta)]\,\mathbf{k}^h_{0})\ \cdots\
      \operatorname{diag}(\tfrac{d}{d\beta}[d^L(-\beta)]\,\mathbf{k}^h_{L})\,\big],
\end{align}
where the derivative of $d^l(-\beta)$ follows from Equations~\eqref{eq: D1D2} and \eqref{eq: dD2}:

\begin{align}
\frac{d}{d\beta}[d^l(-\beta)]
  = -\bigl[D_1^l\, i\, \operatorname{diag}(-l,\ldots,l)\,
     D_2^l(\beta)\bigr]^T.
\end{align}
\section{Astrometry Kernel} \label{ap: astr}
We derive the forward astrometry model using the Wigner-D rotations applied to an astrometric kernel, described above in Appendix \ref{ap: rot}. To derive the astrometric kernel, our approach follows a similar derivation and conventions to those used for photometry measurements in \citet{cowan}. The $ \mathbf{k}^{x}_l$ and $ \mathbf{k}^{y}_l$ vectors for the forward model are the first-moments $k^h_{l,m} = \mathcal{A}_h(Y_l^m(\theta, \phi))$ of each spherical harmonic with no rotation, i.e., inclination $\beta = 0$ corresponding to an equatorial observer and $ t = 0 $, where $\mathbf{k}^{h}_l : h \in \{x, y\}$ denotes the vector of coefficients $ k^{h}_{l,m}$ for $ m = -l $ to $l $.
\begin{align}
k^{x}_{l,m} = \int x_{obs}(\theta, \phi) V(\theta, \phi)  Y_l^m (\theta, \phi) d \Omega, \\
k^{y}_{l,m}  = \int y_{obs}(\theta, \phi) V(\theta, \phi)  Y_l^m(\theta, \phi) d \Omega, 
\end{align}
with $d \Omega = \sin(\theta) d\theta d\phi$ and integration limits $\theta \in [0, \pi]$, $\phi \in [-\pi/2, \pi/2]$ reflecting the visible hemisphere. The astrometric weightings $x_{obs}(\theta, \phi)$ and $y_{obs}(\theta, \phi)$ are the projected x and y-coordinates from the observers perspective. $V(\theta, \phi)$ is the foreshortened visibility, which weights each surface element by the projected area seen by the observer. A point on the surface is visible when the inner product between its coordinates and the observer coordinates is positive:
\begin{align}
\sin(\theta) \sin(\theta_0) \cos(\phi - \phi_0) + \cos(\theta) \cos(\theta_0) \geq 0,
\end{align}
The observer lies along the x-axis facing the equator at position $\theta_0 = \pi/2, \phi_0 = 0$. 
\begin{align}
V_{\beta = 0} = \frac{1}{\pi} \sin(\theta) \cos(\phi)
\end{align}
for $\theta \in [0, \pi], \phi \in [- \pi/2, \pi/2] \; : \; V_{\beta = 0}(\theta, \phi) \geq 0$ is normalized by $\pi$.
For an observer along the x-axis, $x_{obs}$ and $y_{obs}$ correspond to the 3D $y, z$ coordinates. These are $x_{obs} = \sin(\phi) \sin(\theta)$ and $y_{obs} = \cos(\theta)$. Hereinafter, we drop the indices on $x_{obs}$ and $y_{obs}$ and use $x$ and $y$ to refer to the observer's sky-plane coordinates.
\subsection{Deriving the kernel as static first order moments of the spherical harmonics}
The first moment in $x$ is:
\begin{align}
k^{x}_{l,m}  = \frac{1}{\pi}\int \sin(\theta) \sin(\phi) \sin(\theta) \cos(\phi)  Y_l^m(\theta, \phi) d \Omega   \\
k^{x}_{l,m}
= \frac{N_l^m}{\pi}
\underbrace{\int_0^\pi \sin^3\theta\, P_l^m(\cos\theta)\, d\theta}_{I_{u,x}(l,m)}
\;\cdot\;
\underbrace{\int_{-\pi/2}^{\pi/2} \sin\phi\,\cos\phi\, e^{i m \phi}\, d\phi}_{I_{\phi,x}(m)}. \label{eq: k_x}
\end{align}
The second integral over $\phi$ can be expanded as:
\begin{align}
I_{\phi,x}(m) = \int_{-\pi/2}^{\pi/2} \sin(\phi) \cos(\phi) (\cos(m\phi) + i\sin(m\phi)) d\phi  
\end{align}
The first term in this integral is over an anti-symmetric function for all $m$, therefore it is always $0$. The second term is always $0$ for $m=0$, it additionally is $0$ for even $m > 2$.
\begin{equation}
I_{\phi,x}(m)
= i \int_{-\pi/2}^{\pi/2} \sin\phi\,\cos\phi\,\sin(m\phi)\,d\phi
=
\begin{cases}
\dfrac{i \pi}{4}, & m = \pm 2,\\[6pt]
-\dfrac{2i\sin\left(\dfrac{\pi m}{2}\right)}{m^2 - 4}, & m \neq \pm 2 \text{ and $m$ odd}, \\[10pt]
0, & m = 0, m>2 \text{ even}.
\end{cases}
\end{equation}
The first integral in Equation \ref{eq: k_x}, $I_{u, x}(l,m)$, is re-parameterized with $u = cos(\theta)$ and $du = \frac{-d \theta}{sin(\theta)}$:
\begin{align}
I_{u,x}(l,m) =  \int_0^\pi \sin^3(\theta) P_l^m (\cos(\theta)) d \theta  \to \int_{-1}^1 (1 - u^2) P_{l}^m (u) du
\end{align}
Since $1-u^2$ is symmetric over the domain $[-1, 1]$ and the Legendre polynomials $P_l^m$ are antisymmetric when $l-m$ is odd, this integral is $0$ when $l-m$ is odd and non-zero for $l-m$ even. 
Combining these restrictions, $k^{x}_{l,m} $ is $0$ when $l > 2$ is even (for all m), or when $m>2$ is even, also for all $l, m = \pm 1$ and $l,m = \pm 2$.

Similarly we derive the first moment in $y$ as:
\begin{align}
k^{y}_{l,m}
= \frac{N_l^m}{\pi}\,
\underbrace{\int_0^\pi \cos\theta\,\sin^2\theta\, P_l^m(\cos\theta)\, d\theta}_{I_{u,y}(l,m)}
\;\cdot\;
\underbrace{\int_{-\pi/2}^{\pi/2} \cos\phi\, e^{i m \phi}\, d\phi}_{I_{\phi,y}(m)}.
\end{align}
Taking a closer look at the second integral in $\phi$:
\begin{align}
I_{\phi,y}(m) = \int_{- \pi / 2}^{\pi/2} \cos(\phi) (\cos(m \phi) + i \sin(m \phi)) d\phi = \int_{- \pi / 2}^{\pi/2} \cos(\phi) \cos(m\phi)d\phi
\end{align}
This integral evaluates to zero for odd $m > 1$. 
\begin{equation}
I_{\phi,y}(m)
= \int_{-\pi/2}^{\pi/2} \cos\phi\,\cos(m\phi)\,d\phi
=\begin{cases}
\dfrac{\pi}{2}, & m = \pm 1, \\[8pt]
-\dfrac{2\cos\!\left(\dfrac{\pi m}{2}\right)}{\,m^{2}-1\,}, 
    & m\neq \pm 1 \text{ and $m$ even}, \\[10pt]
0, & m>1 \text{ odd}.
\end{cases}
\end{equation}
The first integral $I_{u,y}(l,m)$ is:
\begin{align}
I_{u,y}(l,m) =  \int_0^{ \pi} \cos(\theta)\sin^2(\theta) P_l^m (\cos(\theta)) d \theta \to \int_{-1}^1 u\sqrt{1 - u^2} P_l^m(u) du
\end{align}
The function $u\sqrt{1-u^2}$ is antisymmetric over $[-1, 1]$, and when $l-m$ is even $P_l^m$ is symmetric and the integral is zero.  Therefore $k^{y}_{l,m}$ are zero when $l > 2$ are even or $m$ are odd. The final kernel form is summarized in Equations \ref{eq: k_summary}.
The values of $\mathbf{k}^{h}_l : h \in \{x, y\}$ are integrated numerically with Sympy and shown in Figure \ref{fig: general} for $l \leq 7$.

\subsection{First moments viewed face-on}
The time-dependent first-moment for inclination $\beta = 0$, of a spherical harmonic index by $l,m$ is given by Equation \ref{eq: rot_sph} noting $d_{m,m'}(\beta = 0)  = \delta_{m- m'}$, so that:
\begin{align}
\mu^{x}_{l,m} (t)= e^{im\omega t}k^{x}_{l,m},\\
\mu^{y}_{l,m} (t)= e^{im\omega t}k^{y}_{l,m}.
\end{align}
\subsection{First moments viewed from the pole} \label{ap: pole}
For a pole-on observer ($\beta = \pi/2$), $d^l_{m,m'}(\pi/2)$ is non-zero only for $m = \pm 1$, so the rotated kernel (Equation~\ref{eq: mix}) collapses to a weighted sum over $m'$ for these two values only. We define the rotated kernel weights $p^h_{l,m}$ as the rotated kernel $\mathbf{k}^h_l$ at $\beta = \pi/2$ onto order $m$:
\begin{align}
p^h_{l,m =1} = \sum_{m' = -l}^l d_{1, m'}^l(\beta = \pi/2) k^{h}_{l,m'},
\end{align}
The $m=-1$ weights follow by anti-symmetry of the small Wigner-d matrix: $p^{h}_{l, -1} = -p^h_{l,1}$ for $h \in \{x,y\}$.
\begin{align}
\mu^{x}_{l,m} (t) =
\begin{cases} 
p^{x}_{l,m} e^{i \omega t} & \text{if } l >2 \text{ is odd}, \, m = \pm 1 \\
0 & \text{otherwise},
\end{cases} \\
\mu^{y}_{l,m} (t)  =
\begin{cases} 
p^y_{l,m} e^{i\omega t} & \text{if } l >2 \text{ is odd}, \, m = \pm 1 \\
0 & \text{otherwise},
\end{cases}
\end{align}

\subsection{Photometric kernel} \label{ap: photo}
The photometric kernel is the zeroth-moment analogue of the astrometric kernel: the disk-integrated flux response of each spherical harmonic viewed face-on, with no position weighting. Following the same derivation as above but replacing the first-moment operators $x_{\mathrm{obs}}$, $y_{\mathrm{obs}}$ with unity, the photometric kernel is:
\begin{align} \label{eq: k_photo}
k^{\mathrm{phot}}_{l,m}
= \int V(\theta,\phi)\, Y_l^m(\theta,\phi)\, d\Omega
= \frac{N_l^m}{\pi}\,
\underbrace{\int_{-1}^{1} \sqrt{1-u^2}\; P_l^m(u)\, du}_{I_{u,\mathrm{phot}}(l,m)}
\;\cdot\;
I_{\phi,y}(m),
\end{align}
where $I_{\phi,y}(m)$ is the same azimuthal integral as for the $y$ direction astrometric kernel. This expression for complex spherical harmonics is analogous to Eq. 16 of \citet{cowan} for real spherical harmonics and omitting their time-dependent term $cos(m\phi_0)$. Since $\sqrt{1-u^2}$ is symmetric, the radial integral vanishes when $P_l^m$ is symmetric for odd $l - m$. Additionally $I_{\phi,y}(m) = 0$ for odd $m > 1$. Together, these selection rules give that photometry measures even-degree harmonics: $k^{\mathrm{phot}}_{l,m} = 0$ for all odd $l > 2$. The exception is $l=1$, where $k^{\mathrm{phot}}_{1,\pm 1} \neq 0$.

\bibliography{sample7}{}
\bibliographystyle{aasjournalv7}

\end{document}